\documentclass[9pt,twocolumn,twoside]{pnas-new}

\templatetype{pnasresearcharticle} 

\usepackage{multirow}

\usepackage{natbib}

\usepackage{hyperref}
\usepackage{devanagari}
\usepackage{amsmath}

\begin{document}

\title{Common indicators hurt armed conflict prediction}

\author[a,b,1]{Niraj Kushwaha}
\author[c,d,e]{Woi Sok Oh}
\author[f]{Shlok Shah}
\author[a,1]{Edward D. Lee}

\affil[a]{Complexity Science Hub, Metternichgasse 8, Vienna, 1030, Vienna, Austria}
\affil[b]{Faculty of Physics, University of Vienna, Boltzmanngasse 5, Vienna, 1090, Vienna, Austria}
\affil[c]{Department of Civil, Environmental and Geodetic Engineering, The Ohio State University, 470 Hitchcock Hall, 2070 Neil Avenue, Columbus, 43210
Ohio, USA} 
\affil[d]{High Meadows Environmental Institute, Princeton University, Guyot Hall, Princeton University, Princeton, 08544, New Jersey, USA}
\affil[e]{Department of Ecology \& Evolutionary Biology, Princeton University, Guyot Hall, Princeton, 08544, New Jersey, USA}
\affil[e]{Department of Computer Science, Princeton University, Princeton, 08544, New Jersey, USA}

\leadauthor{Kushwaha}

\significancestatement{Labeling a conflict as, for example, a civil or interstate war presumably gives useful information about eventual size. Surprisingly, we find it does not. To show this, we develop a holistic and vertically integrated methodology that connects highly-resolved data sets on common background indicators with fine-grained conflict data to discover conflict types based on the indicators. We find three overarching conflict types including major unrest, local conflicts, and sporadic and spillover events. Surprisingly, knowing how conflicts fit into one type or another either detracts from predictions of its size or is uninformative. This cautions us about the utility of common data sets for conflict prediction.}

\authorcontributions{Please provide details of author contributions here.}
\authordeclaration{We declare no competing interests.}
\correspondingauthor{\textsuperscript{1}To whom correspondence should be addressed. E-mail: nirajkkushwaha1@gmail.com, edlee@csh.ac.at}


\begin{abstract}
Are big conflicts different from small or medium size conflicts? To answer this question, we leverage fine-grained conflict data, which we map to climate, geography, infrastructure, economics, raw demographics, and demographic composition in Africa. With an unsupervised learning model, we find three overarching conflict types representing ``major unrest,'' ``local conflict,'' and ``sporadic and spillover events.'' Major unrest predominantly propagates around densely populated areas with well-developed infrastructure and flat, riparian geography. Local conflicts are in regions of median population density, are diverse socio-economically and geographically, and are often confined within country borders. Finally, sporadic and spillover conflicts remain small, often in low population density areas, with little infrastructure and poor economic conditions. The three types stratify into a hierarchy of factors that highlights population, infrastructure, economics, and geography, respectively, as the most discriminative indicators. Specifying conflict type negatively impacts the predictability of conflict intensity such as fatalities, conflict duration, and other measures of conflict size. The competitive effect is a general consequence of weak statistical dependence. Hence, we develop an empirical and bottom-up methodology to identify conflict types, knowledge of which can hurt predictability and cautions us about the limited utility of commonly available indicators.
\end{abstract}

\dates{This manuscript was compiled on \today}

\maketitle
\thispagestyle{firststyle}
\ifthenelse{\boolean{shortarticle}}{\ifthenelse{\boolean{singlecolumn}}{\abscontentformatted}{\abscontent}}{}

\firstpage[4]{3}

\dropcap{A}rmed conflicts are multifarious. They span local, civil, and interstate wars, which themselves constitute deeper typologies. Typologies are almost exclusively based on expert assessment of qualitative and political criteria (elaborated on in \textit{Supplementary Information} Appendix~\ref{appendix:heuristic classification}), but an alternative approach that is more reproducible and better formulated for quantitative conflict modeling is to discover conflict types from data. 
The need of a systematic categorization of conflicts has been noted since at least 1989, as emphasized in the \textit{Handbook of War Studies}: ``Although the treatment of war as a generic category has proven useful until now, future research may require the systematic delineation among several categories, each of which may require a separate theoretical treatment"~\cite{midlarskyHandbookWarStudies2011}. This is now feasible in the modern era of conflict data and computational advances. 

A data-oriented approach is appealing when considering the manifold drivers of conflict. A broad literature identifies potential drivers like climate, especially deviations from historical norms \cite{machClimateRiskFactor2019, ayanaExaminingRelationshipEnvironmental2016, hsiangQuantifyingInfluenceClimate2013}, economic development \cite{collierGreedGrievance2004,fearonEthnicityInsurgency2003}, and infrastructure \cite{taoHybridApproachModeling2016}. While not necessarily direct drivers, proxies can capture the effects of drivers such as geographic features \cite{wollebSharedRiversInterstate2000, braithwaiteGeographicSpreadMilitarized2006, buhaugGeographyCivil2002}, raw demographics \cite{raleighPopulationSizeConcentration2009}, and demographic composition \cite{brunborgDemographyConflictViolence2005, ismailWhyYouthParticipate2021, williamsMicroLevelEventCenteredApproach2012, bohra-mishraIndividualDecisionsMigrate2011}. 
Each mentioned category alone constitutes a multi-dimensional feature space, such as how geography includes elevation and distance from water bodies. Complicating this picture further, drivers do not act independently, but affect each other in feedback and feedforward loops, forming a complex, interdependent network. Thus, the set of possible drivers and proxies thereof specifies a combinatorially large space of possible interactions that makes it difficult to represent with a simple organizational framework. 
We might picture this problem as a high-dimensional Cartesian space, where each axis represents the state of a conflict driver. Densely populated regions of the space indicate where many different conflict tend to manifest similar properties. If pairs of drivers move together or against each other, we would naturally expect a multi-peaked probability density, or a rugged landscape \cite{hopfieldNeuralNetworksPhysical1982}. The peaks in the density would represent global states in which the full combination of driver states encodes archetypal conflicts. 
In complement, the valleys highlight combinations of drivers antithetical to conflict formation. Such a map would show how conflict drivers coalesce into a reduced set of conflict archetypes.

Inspired by this picture, we assemble a high-dimensional representation of conflicts by combining detailed spatiotemporal information, including publicly available conflict data set, the Armed Conflict \& Event Data Project (ACLED), and background indicators that may inform about conflict properties. The aggregated list of properties allows us to enumerate hypothesized drivers and their proxies in a way that captures a pseudo-representation of (and would be extendable to) a more complete representation. We build an unsupervised learning approach to search this space for peaks in probability density, and we show that the clusters of conflicts reduce to a minimal description, captured in three peaks that represent interpretable conflict categories. Finally and surprisingly, we show that such conflict categorization does not inform conflict size and therefore is in direct competition with its prediction, a sobering reminder about the limits to publicly available information for gaining predictive insight.

\begin{figure}[!t]
    \centering
    \includegraphics[width=\linewidth]{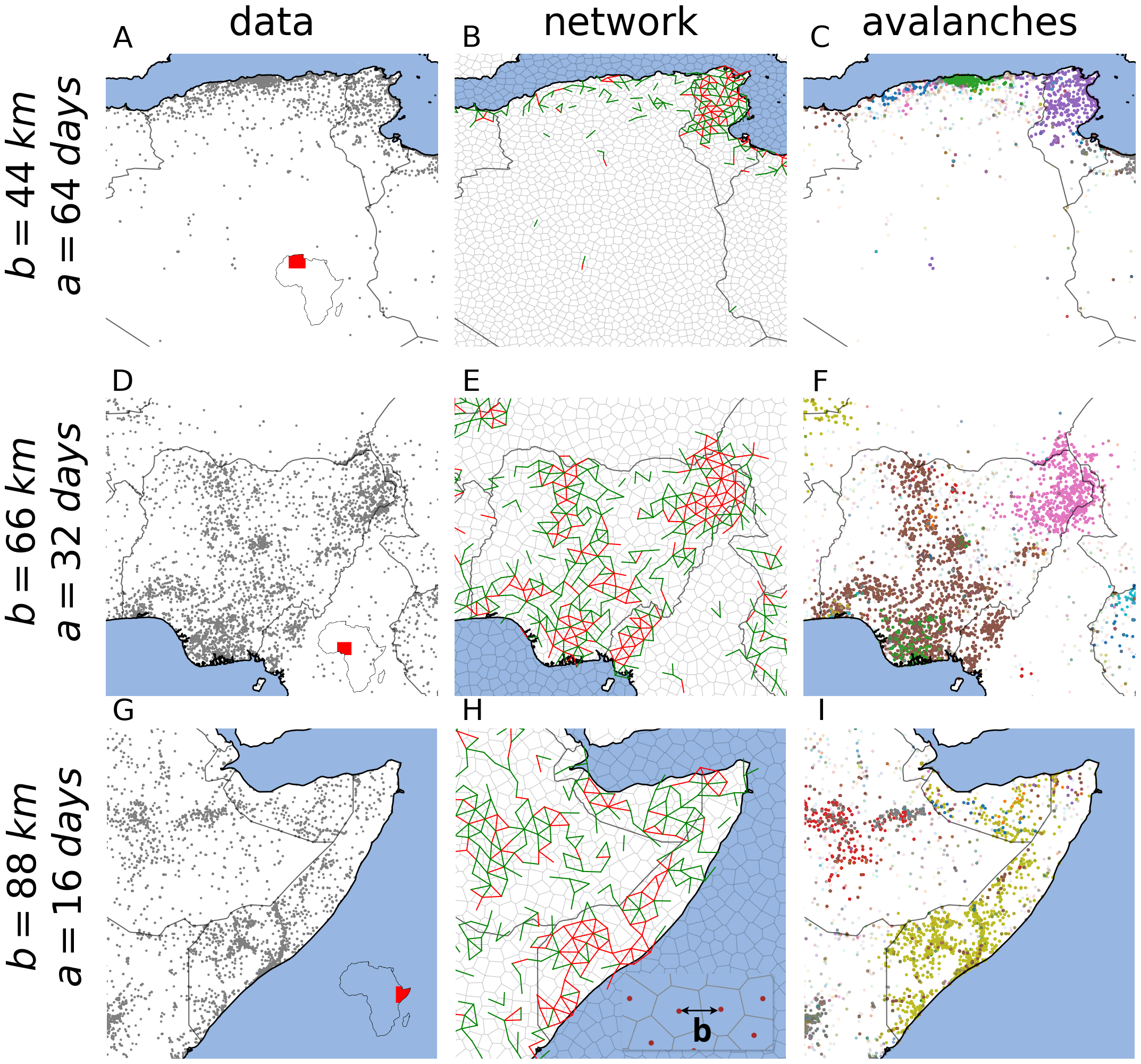}
    \caption{Conflict avalanches are generated from (A, D, G) disaggregated conflict event data from ACLED shown for Algeria, Nigeria and Somalia. Each point represents a conflict event at a specific location and time. Geographic area is divided into pseudorandom Voronoi spatial bins of size $b$ kilometers, and the time series is segmented into temporal bins of $a$ days. We then infer a (B, E, H) network by calculating directed transfer entropy for pairs of spatial bins. Red links are bidirectional, while green are unidirectional. Conflict avalanches (C, F, I), defined as sequences of conflict events connected via the transfer entropy network. Each color in C, F, and I correspond to a different conflict avalanche. Conflict events that belong to avalanches with fewer than 50 events are in grey. Inset in H shows Voronoi grids with their centers and the distance $b$ between these centers. For details on conflict avalanche generation see reference \citenum{kushwahaDiscoveringMesoscaleChains2023a}.}
    \label{fig:avalanches}
\end{figure}

We rely on ACLED, the largest, publicly available conflict database that includes about ${\sim}10^6$ conflict events between 1997-2024 that are largely collected from news reports in coordination with local partners \cite{raleighIntroducingACLEDArmed2010}. Each event in the database represents an instance of conflict at a particular coordinate, on a particular day, with purported measures of fatalities or involved actors. We show the spatial distribution of events in particular regions in Figure~\ref{fig:avalanches}A, D, and G. We focus on Africa, the largest contiguous landmass and with the most extensively reported data. Data sets like ACLED are valuable because they provide a fine-grained view into conflicts, but they pose a complementary difficulty: conflict events do not happen independently of one another, so it is useful to first group the events into chains of related activity such as battles or wars. Common techniques for grouping events use administrative boundaries like country borders \cite{gutierrez-romeroConflictsIncreased2022} or combine events with the same purported actors \cite{dowdCulturalReligiousDemography2015}. The heuristic techniques, however, do not leverage statistical patterns in the timing and location of conflict activity.

To account for statistical relationships in the observed dynamics, we take neighboring geographic regions and compute directed links of time-lagged predictability between them as we diagram in Figure~\ref{fig:avalanches}. We first define a distance over which we look for such relationships in time and geographic space, defining a resolution time $a$\,days and distance $b$\,km. Operationally, we subdivide Africa into a pseudorandom Voronoi lattice with regions of length scale $b$\,km and coarse-grain time into bins of length $a$\,days. As a result, we have a pattern of conflict activity in any particular cell at a time indicating when there are conflict events detected or not. We then search for statistical dependence between adjacent cells by asking whether or not activity in the adjacent cell helps predict better activity in the target cell. The quantity that measures this gain in predictability is the transfer entropy, a generalization of Granger causality that accounts for nonlinear dependence \cite{schreiberMeasuringInformationTransfer2000}. By collecting pairs of adjacent cells that show significant transfer entropy, we obtain a directed network that indicates paths along which conflict activity is temporally predictable as in Figure~\ref{fig:avalanches}. We then construct chains of conflict events by grouping together events that have ocurred simultaneously according to the given resolution $b$ and $a$ or in any adjacent site at a sequential time to which there is an outgoing path. These chains of conflict are \textit{conflict avalanches} (for more details see reference \citenum{kushwahaDiscoveringMesoscaleChains2023a}). In a mesoscale between $b \approx 60$~km and $b \approx 400$\,km and $a \approx 4$~days to $a \approx 128$~days, conflict avalanches in aggregate display cascades of activity with nontrivial, long-range correlations and align with mechanism identified in field studies. We show examples of conflict avalanches thus recovered in Figures~\ref{fig:avalanches}C, F, and I.  Importantly, conflict avalanches are only constructed from patterns of activity, without explicit use of detailed information directly involving conflicts (e.g., actors, fatalities, etc.), allowing us to then explore how these additional features distinguish avalanches from one another.

\begin{figure*}[!t]\centering
    \includegraphics[width=\linewidth]{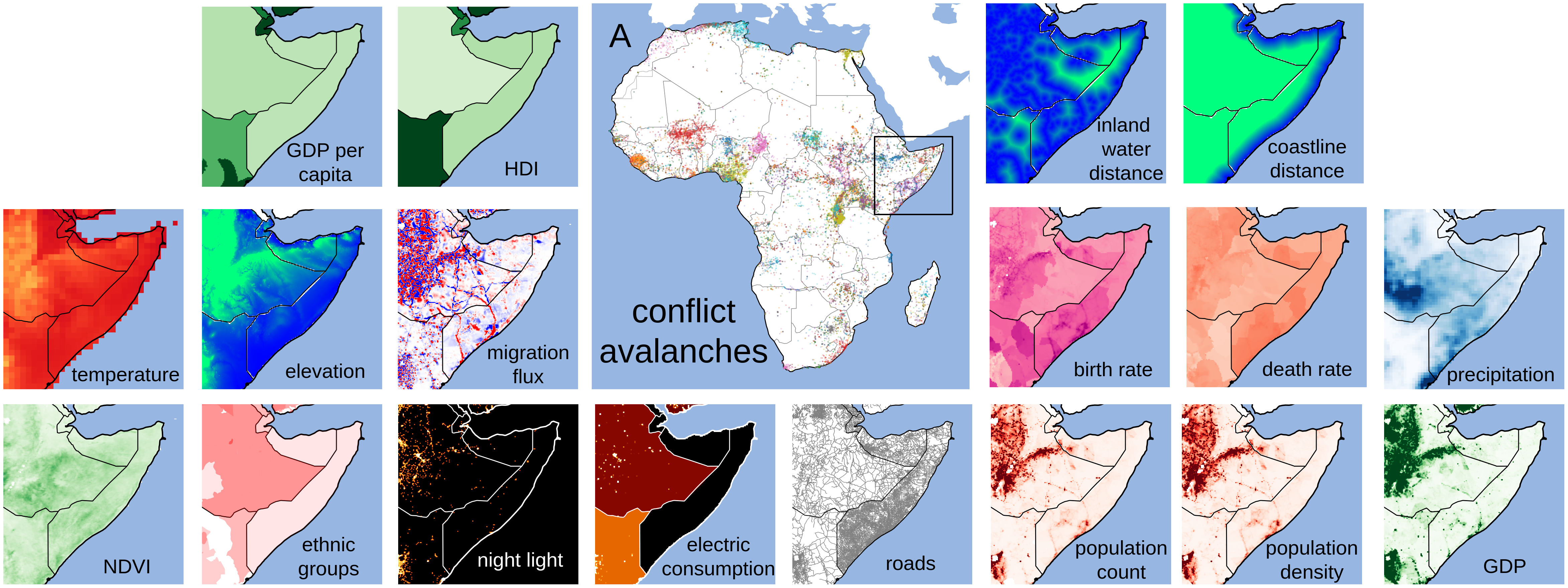}
    \caption{Datasets. A) Disaggregated conflict data from ACLED. Each point is an individual conflict event. These are grouped into conflict avalanches denoted by color ($b\approx66$\,km and $a=30$\,days). Remaining panels showcase background indicators (from left to right and top to bottom: GDP per capita, HDI, inland water distance, coastline distance, temperature, elevation, migration flux, birth rate, death rate, precipitation, NDVI, ethnic groups, night light, electric consumption, roads, population count, population density and GDP).}\label{fig:data}
\end{figure*}

For each conflict avalanche thus obtained, we assemble a set of factors associated with armed conflict. These factors largely fall into six major categories that are usually considered separately in the literature: climate \cite{machClimateRiskFactor2019, ayanaExaminingRelationshipEnvironmental2016}, economics \cite{collierGreedGrievance2004, fearonEthnicityInsurgency2003}, geography \cite{wollebSharedRiversInterstate2000, buhaugGeographyCivil2002, braithwaiteGeographicSpreadMilitarized2006}, infrastructure \cite{taoHybridApproachModeling2016, zhukovRoadsDiffusion2012}, raw demographics \cite{thalheimerLargeWeatherConflict2023}, and demographic composition \cite{brunborgDemographyConflictViolence2005a, ismailWhyYouthParticipate2021, williamsMicroLevelEventCenteredApproach2012, bohra-mishraIndividualDecisionsMigrate2011}. 
\textit{Climate}, often associated with increase in resource strain and probability of onset of armed conflicts \cite{shivaClimaticConditionsInternal2022}, includes rising temperatures (linked to an increased risk of conflict and the persistence of ongoing conflicts in Africa \cite{vanweezelLocalWarmingViolent2020}), variation in precipitation (associated with communal conflicts in Ethiopia and Kenya \cite{vanweezelClimateConflictPrecipitation2019}), and the Normalized Difference Vegetation Index, so-called NDVI (observed to increase in Afghanistan in areas affected by armed conflict, possibly due to human migration that reduces anthropogenic pressures on the environment \cite{zhangImpactArmedConflict2023}).
\textit{Economics} is frequently studied to assess the onset and impacts of armed conflict. This includes the Human Development Index, or HDI (a proxy for the widely discussed detrimental effects of wars on human development \cite{anandHumanDevelopment1994, vescoImpactsArmed2025}), as well as GDP and GDP per capita (the most common proxies for economic prosperity and are used to estimate the economic cost of armed conflict to a country \cite{lindgren2004measuring, lopezEconomicImpactArmed2005}). 
\textit{Geography}, often recognized as significant in influencing the dynamics and spread of armed conflicts, includes proximity to water bodies, which has become a common part of the political rhetoric in the context of conflicts since as early as 1967 \cite{wollebSharedRiversInterstate2000} and elevation, which has been hypothesized to shape conflicts by influencing actions and motivations of armed groups \cite{linkeMountainousTerrainCivil2017}. 
\textit{Infrastructure}, widely regarded as critical in determining strategic areas, includes distance from roads \cite{raleighSeeingForestTrees2010}, electric consumption, which has been shown to decline during times of crisis in Syria \cite{alhajomarEnergyPovertyFace2023}, and mobile phone coverage, where an increase has been linked to a higher probability of conflict occurrence in Africa \cite{ackermannMobilePhoneCoverage2021}.
\textit{Raw demographics} like population count and density \cite{raleighSeeingForestTrees2010} are linked to increase in likelihood of armed conflict due to resource constraints and governance challenges \cite{raleighPopulationSizeConcentration2009}.
In contrast, \textit{demographic composition} includes net migration (a major factor cited in relation to conflict and a long-standing international policy concern \cite{williamsHowArmedConflict2021}), ethnic diversity (linked to conflicts from competing ethnonationalist claims to power and still a major aspect of study \cite{cedermanWhyEthnicGroups2010}), birth and death rates (directly affected by an ongoing armed conflict and association has been shown between conflict and higher rates of child and maternal mortality in sub-saharan Africa \cite{ohareFirstNoHarm2007}). While similar to raw demographics, it is often considered separately. 
\textit{Supplementary Information} Appendix~\ref{appendix:data} gives a dataset summary and further specifications. The set excludes some commonly cited factors like infant mortality and the Gini coefficient since only a handful of datasets cover every African country in high-resolution and are updated at least annually (except unchanging features like geography), criteria that limit us to the period 2000-2015. Therefore, in total 22 datasets, belonging to 6 variable categories, were collected adhering to the data quality constraints as shown in Figure~\ref{fig:data}.

All together, we have for each conflict avalanche a detailed profile for each event indexed $i$, or the vector $\vec{e}_i$, whose 22 dimensions form six major categories. In its full complexity, this is a partially ordered set $c_j=\{\vec{e}_i\}_j$ for each avalanche $j$ whose size varies with the number of events in the avalanche. The events in any given avalanche, however, are largely redundant because many are similar to each other (see \textit{Supplementary Information} Figure~\ref{appendixFig:within_std}). Furthermore, their raw values obscure the fact that fluctuations away from either historical or geographic tendencies of conflict regions is most relevant to conflict \cite{wischnathClimateVariabilityCivil2014, vestbyWhyPoorCountries2021, carterPlacesHideTerrain2019}. We would like to compress the representation to squeeze out redundancy and to highlight fluctuations away from the typical value.

\begin{figure}[!t]
    \centering
    \includegraphics[width=\linewidth]{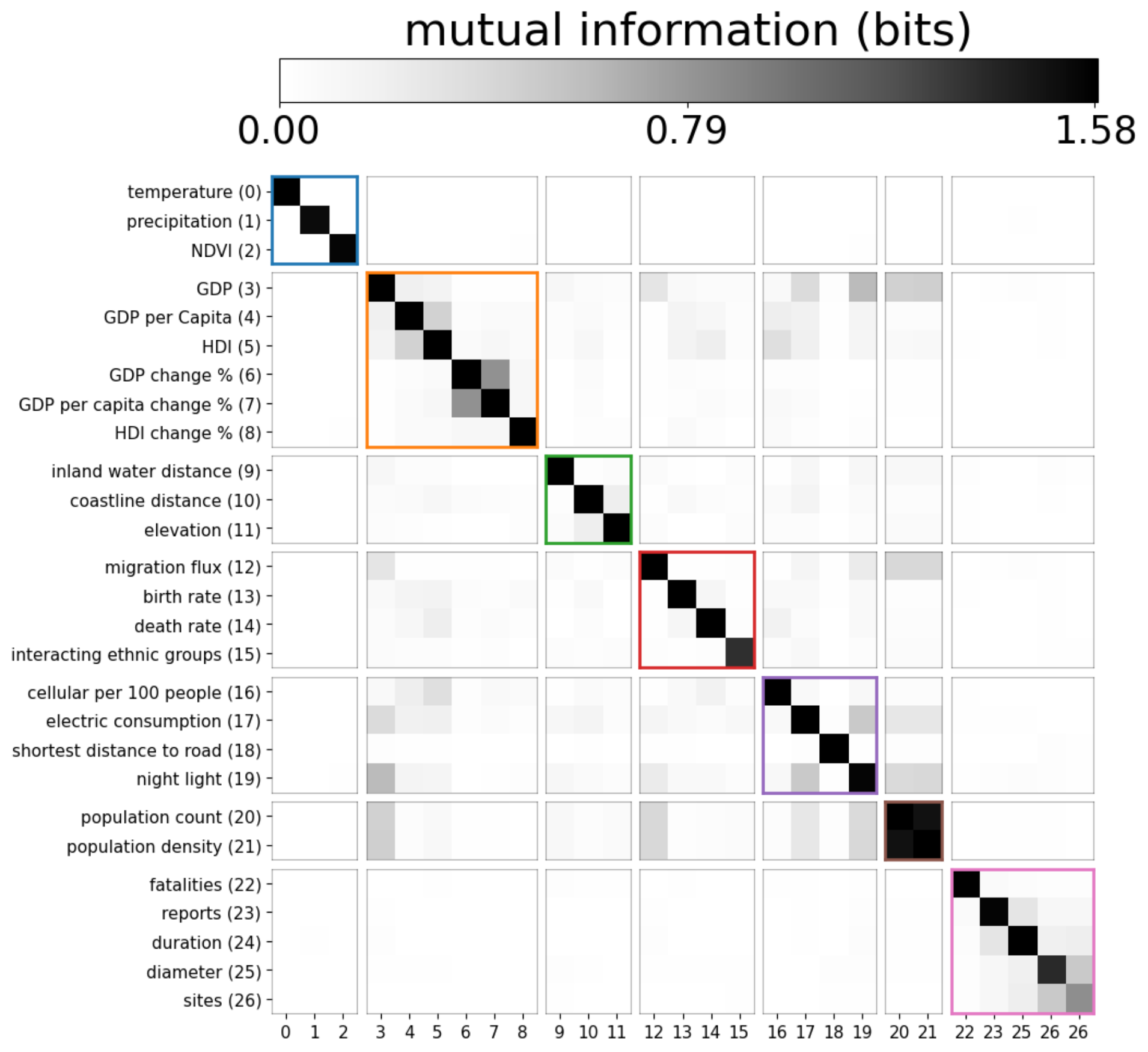}
    \caption{Mutual information matrix for pairs of background indicators used as conflict variables. Diagonal entries indicate the entropies as estimated with the Nemenman-Shafee-Bialek (NSB) estimator \cite{nemenmanEntropyInference2002}.}
    \label{fig:MI}
\end{figure}

To develop such a procedure, we treat separately climatic and non-climatic variables. This is because changes in climatic variables are most meaningful in relation to historical values at that region, whereas fluctuations in non-climatic variables like GDP are most meaningful when compared with other conflict-prone regions. The resulting coarse-graining procedure (as discussed further in \textit{Supplementary Information} Appendix \ref{appendix:conflict_vector}) first transforms the variables into deviations about the median, labels the deviations as unusually positive or negative relative to the median by an equipartition of percentile rank, and reduces each variable into the typical deviation for the avalanche. When we choose $L=3$ divisions, for example, the procedure results in a vector $\vec{c}_j$ for avalanche $j$, where the value of the $k$th climatic variable whose mode is below the 33rd percentile is assigned $c_{jk}=-1$, between the 33rd and 67th percentiles $c_{jk}=0$, and the remainder $c_{jk}=1$. Similarly for non-climatic variables $k$th variable whose value is below the 33rd percentile is assigned $c_{jk}=-1$, between the 33rd and 67th percentiles $c_{jk}=0$, and the remainder $c_{jk}=1$. We focus on $L=3$ as the simplest representation of background properties that distinguishes extreme variation away from the median, but our results do not depend on this choice (see \textit{Supplementary Information} Figures~\ref{fig:PCA_SI}B, C, and D for more details). After these steps, the feature space now consists of 22 dimensions, populated with 5,659 avalanches having an entropy of $S \approx 17.7$ bits\footnote{The entropy is calculated using the NSB estimator \cite{nemenmanEntropyInference2002}. The elements of these vectors correspond to deviations from the median for each variable, resulting in a total discrete state space of size $3^{22}$.} out of a state space of size $\sim 10^{5}$. The large estimated entropy reflects the diversity of avalanches. Having undertaken this procedure, we obtain for each avalanche a vector of conflict descriptors as a ternary code that represents how extreme or median the typical value of each variable is in comparison to history or to contemporary, geographic peers.

As an overview of resulting feature vectors, we show the correlation structure between the properties in Figure~\ref{fig:MI} for a representative choice of separation scales $b\approx 66$\,km and $a=30$\,days. We show the mutual information $I[X;Y]$, a nonlinear measure of dependence between two random variables $X$ and $Y$ with joint probability distribution $p(x,y)$,

\begin{align}
I[X;Y] = \sum_{x\in X, y\in Y} p(x,y) \log \left(\frac{p(x,y)}{p(x)p(y)}\right).
\end{align}

The mutual information is zero when two variables are uncorrelated, $p(x,y)=p(x)p(y)$. While some of the variables are strongly correlated because they involve combinations of the same underlying variables, we also find that there is little information between the variable categories, indicating that they present largely independent measures of the background on which conflict evolves.

The weak correlation structure, large entropy, along with the high-dimensional space implies that feature vectors are mostly equidistant from one another, suggesting that the exact string of variable values leads to a sparsely populated feature space and would fail to highlight similar avalanches. Instead, a simplification that still captures how extreme (or median) the variables corresponding to avalanches are is a ``bag-of-words'' representation counting the total number variables below $n_{-1}$, at $n_0$, and above $n_1$ median within each variable category. This category separation makes sure that we distinguish between the types of extremity that correspond to each of the six variable categories. The resulting mutual information matrix again highlights the relatively weak correlations between variable categories (see \textit{Supplementary Information} Figure~\ref{appendixFig:MI_SI}). Thus, the feature space now consists of 18 dimensions, including 12 free dimensions and 6 given by a normalization constraint for each variable category.

A simple model that accounts for each of the six variable categories and their respective counts is the product of six multinomials, a multi-multinomial.\footnote{Note that while this assumes independence along the variable categories for any given conflict type, it can recover correlations between the variable categories across the multiple centroids found, once fitted to the data.} When each is indexed $\nu$, the probability of any particular observation of counts for a given avalanche with $n^\nu = n^\nu_{-1}+n^\nu_0+n^\nu_1$ is

\begin{align}
M_\theta(n) &=  \prod_{\nu=1}^6 \left(\frac{n^\nu!}{\prod_{j=-1}^1 n^\nu_j!} \prod_{j=1}^M \theta_j^{n_j^\nu} \right)\label{eq:M3}
\end{align}

\begin{align}
\sum_{j=-1}^1 \theta_j^{n_j^\nu} &= 1.
\end{align}
Finally, a single multi-multinomial $M_\theta$ represents only one type of conflict avalanches, meaning that to capture several we define a mixture of $K$ multi-multinomials indexed $i$, normalized weight $\pi_i$, and different parameter sets $\theta_i$, 
\begin{align}
p(\vec x|\theta) &= \sum_{i=1}^K \pi_i M_{\theta_i}. \label{eq:m4}
\end{align}
Eq~\ref{eq:m4} is variation on a ``bag of words'' model, where a bag holds a mix of three different ``words,'' and there is a separate bag for each of the six variable categories. In contrast to a ``single bag'' model---a multinomial mixture model (M3), which is widely used for unsupervised clustering---we use the multi-multinomial mixture model (M4) to search for the peaks in the distribution that represent clusters of similar conflict avalanches.

To solve for the parameters, we find the maximum likelihood estimator of M4 given the data using Eq~\ref{eq:m4}. This can be done with the expectation-maximization algorithm, which reduces to alternating between finding the centroid of the points that belong in the cluster and associating avalanches with the nearest centroid until convergence (see \textit{Supplementary Information} Appendix~\ref{appendix:m4} for the modified derivation for M4). We iterate for $10^3$ random initial starting conditions and take the best result, obtaining $K$ avalanche classes, where the set of solutions $\theta^*$ indicates local peaks in the probability distribution. We take the hard clustering limit to assign each avalanche to its most likely cluster according to the maximum value of $\tau_{ij}$, the probability that an avalanche $i$ in cluster $j$, as long as $\tau_{ij}\geq1/2$. Otherwise, it is not assigned to a cluster, but these constitute a small minority of 6\% when $K\leq15$ (see \textit{Supplementary Information} Figure~\ref{appendixFig:not_hard_clustered}). We set this as our threshold to determine the upper limit of $K$ in the further analysis. At the end of the procedure, we have $K$ localized clusters to which we have assigned the great majority of avalanches and thus have identified the peaks in the avalanche feature space that we had set out to find.

\begin{figure}[!t]\centering
    \includegraphics[width=\linewidth]{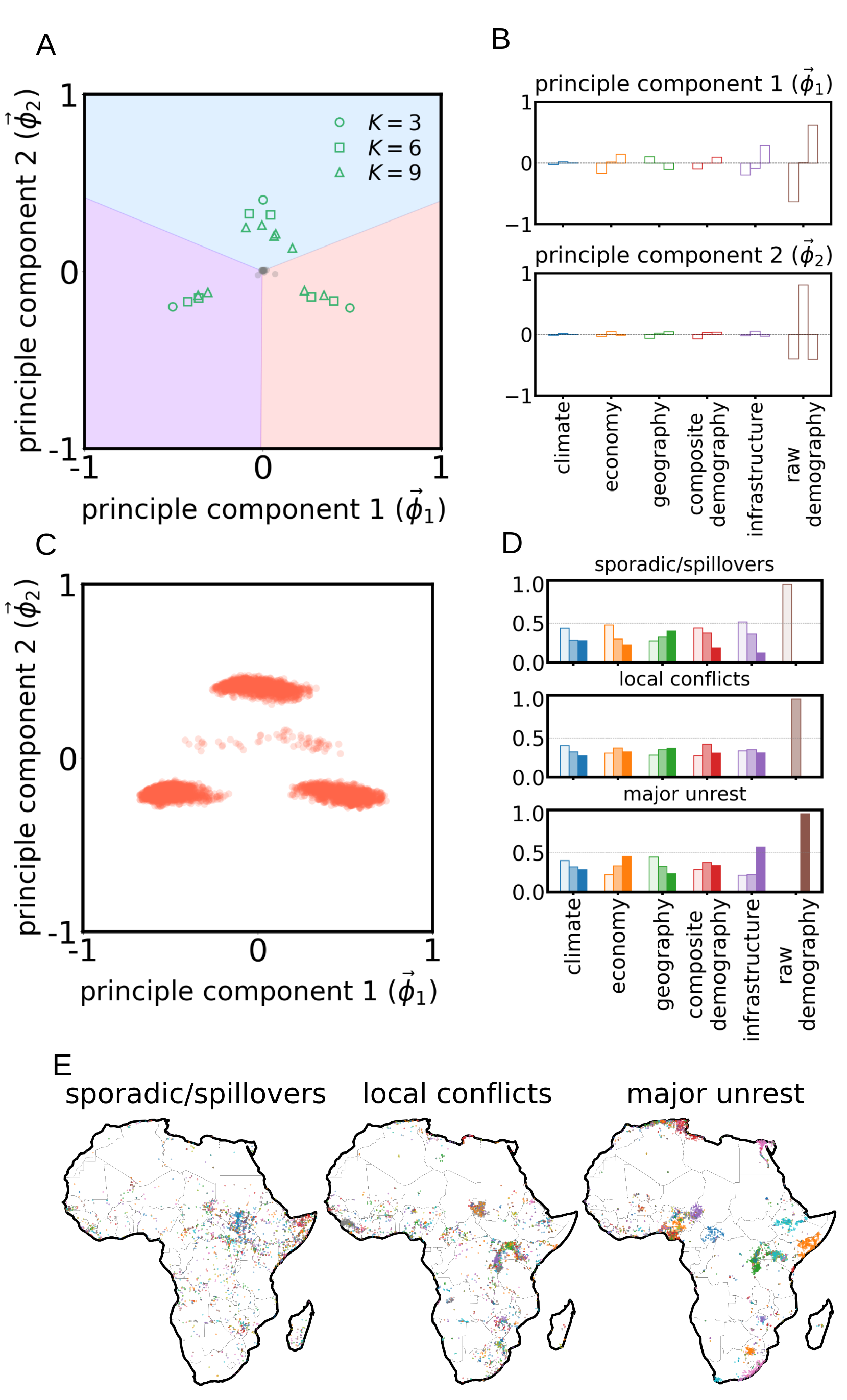}
    \caption{Three armed conflict archetypes. A) Cluster centroids (for $K=3,6,9$) projected onto the first two principal components $\vec\phi_1$ and $\vec\phi_2$ from $K=3$. Bisectors demarcate the three archetypes, where red is major unrest, blue local conflict, and purple sporadic and spillover. Grey points show centroids from shuffled nulls. B) Corresponding eigenvectors $\vec\phi_1$ and $\vec\phi_2$. C) Projection of each conflict avalanche onto eigenvectors. D) Centroids for clusters at $K=3$. Bars are grouped into sets of three corresponding  values of $\theta_i$ that give the frequency of below, within, and above median properties. Analysis done for conflict avalanches generated at scales $b\approx66$\,km and $a=30$\,days, but are representative. See \textit{Supplementary Information} Figure~\ref{appendixFig:PCA_SI_2} for other scales. For a schematic overview of our methodology see \textit{Supplementary Information} Figure~\ref{appendixFig:schematic}.}\label{fig:proj}
\end{figure}

The number of clusters $K$, however, is an important hyperparameter; it determines whether we overlook important peaks in the feature spaces or overfit the distribution in the limit of large $K$. Notably, we find that the cluster centroids consistently separate into three superclasses as we increase $K$. To see this, we start with parameters for $K=3$ and project them into the first two principal components of the covariance matrix $C$ denoted $\vec \phi_1$ and $\vec \phi_2$---the element $C_{ij}$ is the covariance of parameter set $\theta_i$ with $\theta_j$. The normalization condition for $\theta$ stipulates in the nontrivial case that we obtain three centroids. This points to a tripartite division of parameter space, and we correspondingly color the plane purple, blue, and red by bisecting the centroids in Figure~\ref{fig:proj} (the shown bisectors are projections down from the full space). For $K>3$, we project the solved model parameters onto the same eigenvectors. Yet, we again find that new clusters recover the same tripartite structure, even up to the fine-grained case $K=15$ (see \textit{Supplementary Information} Figure~\ref{fig:PCA_SI}A for $2<K<16$ and movie in reference \citenum{kushwahaClusterCentroids2025}). Furthermore, we find that the first two dimensions capture substantial variation in the parameters because the total variance captured remains above $>65\%$ (\textit{Supplementary Information} Figure~\ref{appendixFig:variance}). As a final check, we generate shuffled versions of the data set, where the variable values are randomly swapped between avalanches, thus destroying any non-trivial patterns between variables and between variable categories. The resulting eigenvectors and clusters show relatively flat $\pi_i$ distributions with high entropy $S=-\sum_i \pi_i \log \pi_i$ and precipitate exclusively at the origin as the stain of gray points shows in Figure~\ref{fig:proj}A. Furthermore, the result is not a trivial outcome of having set $L=3$; it depends weakly on the degree of quantization and we tested up to senary variables (\textit{Supplementary Information} Figure~\ref{fig:PCA_SI}B, C, and D). These lines of evidence all point to the conclusion that the triangle in Figure~\ref{fig:proj}A is preserved regardless of the number of clusters that we seek out, is not replicated under null model with shuffled avalanche vectors, and is not a trivial result of the ternary coding for the variables.

Each of the three corners in Figure~\ref{fig:proj}A reveals a distinct conflict archetype. At the bottom right, avalanches exhibit extensive spread, frequently traversing national borders and often persisting for years up to decades. Notable examples within this cluster include the Al-Shabaab insurgency \cite{andersonUnderstandingAlShabaabClan2015}, Boko Haram insurgency \cite{walkerWhatBokoHaram2012}, and the Central African Republic Civil War \cite{arieff2014crisis}. Given their geographic impact and prolonged duration, we name these \textit{major unrest}. In contrast, conflicts within the clusters at the top are better localized, typically confined within national borders, and tend to be shorter, generally lasting from a few months to a year. Examples of conflicts in this cluster include the conflicts in Ituri \cite{faheyTroubleIturi2011} and Kivu \cite{ekyambaAssessingChallengesArmed2022}, the Seleka and anti-Balaka conflict \cite{kah2014anti,isaacs-martinSelekaAntiBalakaRebel2017}, and local clan violence in Somalia between 2003 and 2004. We name these \textit{local conflicts}. Lastly, the clusters at the bottom left encompass minor and sporadic conflicts that are small and brief. The cluster includes spillover conflicts, such as extensions of the Al-Shabaab insurgency across Somalia, including the conflict around Bosaso \cite{SomaliaAlShabaabAttack2017}. We name these \textit{sporadic and spillover events}. The three vertices constitute a triangle of madness.

The triangle shown in Figure~\ref{fig:proj}A is for separation distance $b \approx 66$\,km and time $a=30$\,days, but it is preserved when we change $b$ and $a$ to obtain bigger or smaller conflict avalanches (\textit{Supplementary Information} Figure~\ref{appendixFig:PCA_SI_2}). This holds all the way down to event-level data $b\sim 1$\,km and up to the largest scales $b\sim10^3$\,km. For $a$, the range includes 1\,day to 128\,days, although the self-similarity may be less surprising given some of the data sets only change annually. The consistency indicates that the coarse-graining procedure for obtaining conflict avalanche features preserves the topology of the probability distribution encoded at the event-level, or that the peaks in the density are fixed. As we generate avalanches by joining events by geographic proximity along the transfer entropy graph (see Supplementary material of reference \citenum{kushwahaDiscoveringMesoscaleChains2023a}), proximate events must be more similar to one another than to the mean conflict event to preserve avalanche properties. For some of the variables, this is a result of data resolution (we have only country-level resolution for cellular phones per 100 people, so by definition proximate events are similar), but for majority of variables employed in our procedure, this self-similarity is more surprising, especially at the range of $\sim 400$\,km. Self-similarity is not a central-limit-type phenomenon and supports observations of scaling in conflicts \cite{bohorquezCommonEcology2009,clausetGeneralizedAggregationDisintegration2010,johnsonSimpleMathematical2013, leeScalingTheoryArmedconflict2020}.  
Thus, the preservation of the triangle of madness across scales of resolution reiterates the importance of geographic proximity and the patterns of self-similarity \cite{leeScalingTheoryArmedconflict2020} encoded in it.

\begin{figure}[!t]\centering
    \includegraphics[width=\linewidth]{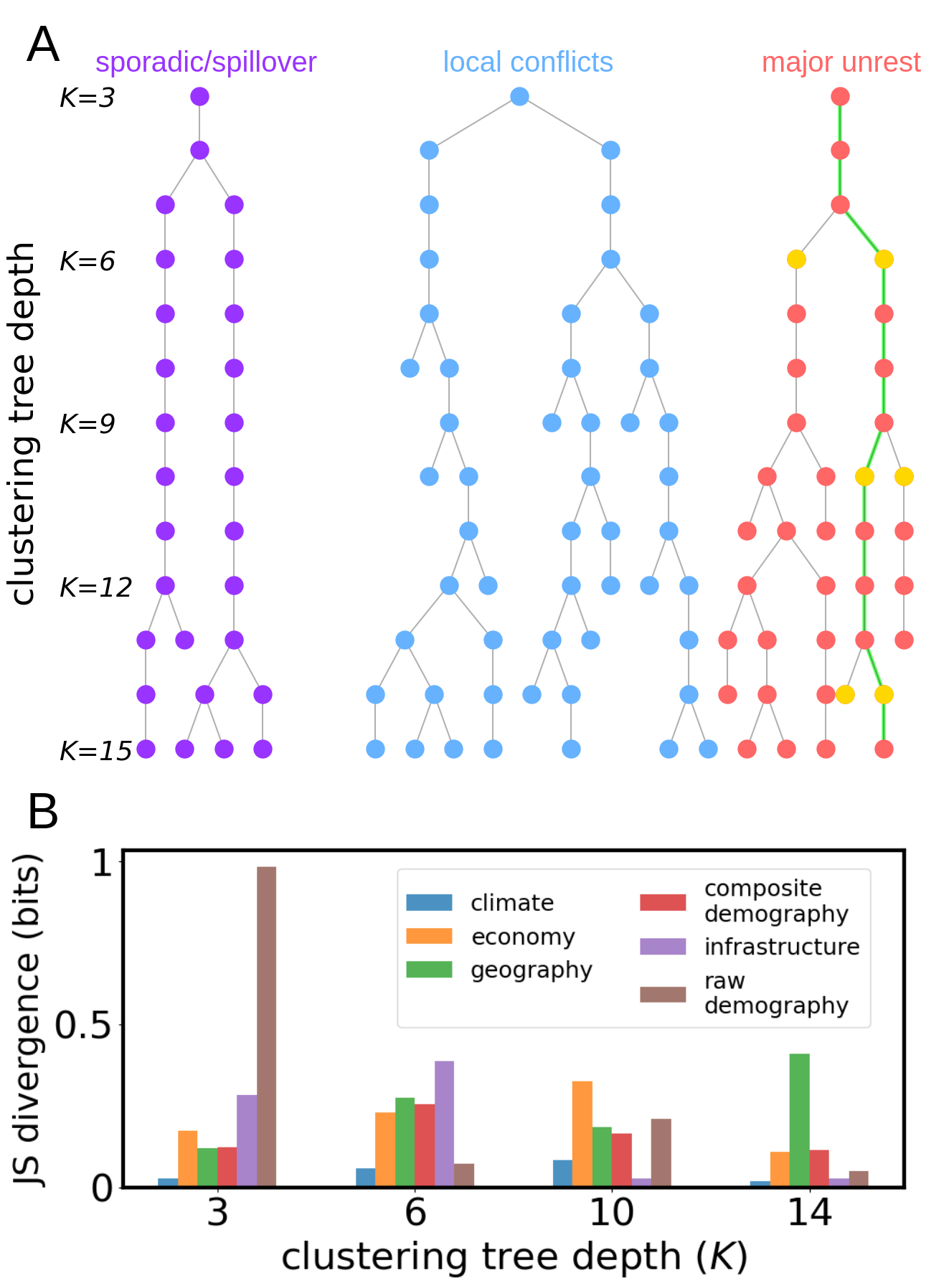}
    \caption{(A) Reconstructed conflict taxonomy. (B) Discriminative variable categories at branching points of the tree as highlighted by JS divergence. This represents only the green branch in panel A. Along this branch, clusters split at $K=6$, $K=10$, and $K=14$, as highlighted in yellow. At the splits, the most discriminative variable categories raw demography, infrastructure, economics, and geography. Same ordering is observed across other branches of the tree (see \textit{Supplementary Information} Figure~\ref{appendixFig:JS} for other branches). Taxonomic tree with parameter values and avalanches of each cluster are shown in \textit{Supplementary Information} Figures~\ref{appendixFig:thetas} and \ref{appendixFig:clusters}.}\label{fig:hierarchy}
\end{figure}

While preserving the overall tripartite arrangement, larger cluster number $K$ leads to a hierarchy of conflict types that form a taxonomy. At the highest level, the taxonomy shows the three main branches that we describe above, and this is well-determined by raw demographics as shown by values of $\theta$ in Figure~\ref{fig:proj}D. Upon increasing $K$, we obtain a finer-grained description. We show a depiction of the inferred taxonomy in Figure~\ref{fig:hierarchy}. As we increase $K$, we consider the new centroids that are found and associate them to the closest centroid at $K-1$  by the Jensen-Shannon distance. In principle, there is no guarantee that the clusters at larger $K$ are similar to the ones at smaller $K$. We find, however, that the resulting clusters are either almost the same as or a split of one of the clusters at $K-1$, just as we would expect for a hierarchical taxonomy (see \textit{Supplementary Information} Figure \ref{appendixFig:thetas} and \ref{appendixFig:clusters} for a detailed look). At each split in Figure~\ref{fig:hierarchy}A, we depict a new branching, and there we can measure which variable categories best distinguish the new subtypes, again using the Jensen-Shannon divergence. Tracing each branch down, we find that the most common pattern of feature importance with each new subdivisions in ranked order is raw demographics, infrastructure, economy and geography. This consistency reveals them to be the most discriminative indicators of conflict type.

\footnotetext{The Jensen-Shannon (JS) divergence is a symmetric measure that quantifies the similarity between two probability distributions \( P \) and \( Q \), defined as:
\[
JS(P || Q) = \frac{1}{2} \sum_{x \in \mathcal{X}} P(x) \log_2 \frac{P(x)}{M(x)} + \frac{1}{2} \sum_{x \in \mathcal{X}} Q(x) \log_2 \frac{Q(x)}{M(x)}
\]
where, \( M = (P + Q)/2 \) is the midpoint (or mixture) distribution. JS divergence is bounded by $0$, for identical distributions, and $1$ for disjoint distributions.}

\begin{figure}[!t]
\centering
\includegraphics[width=\linewidth]{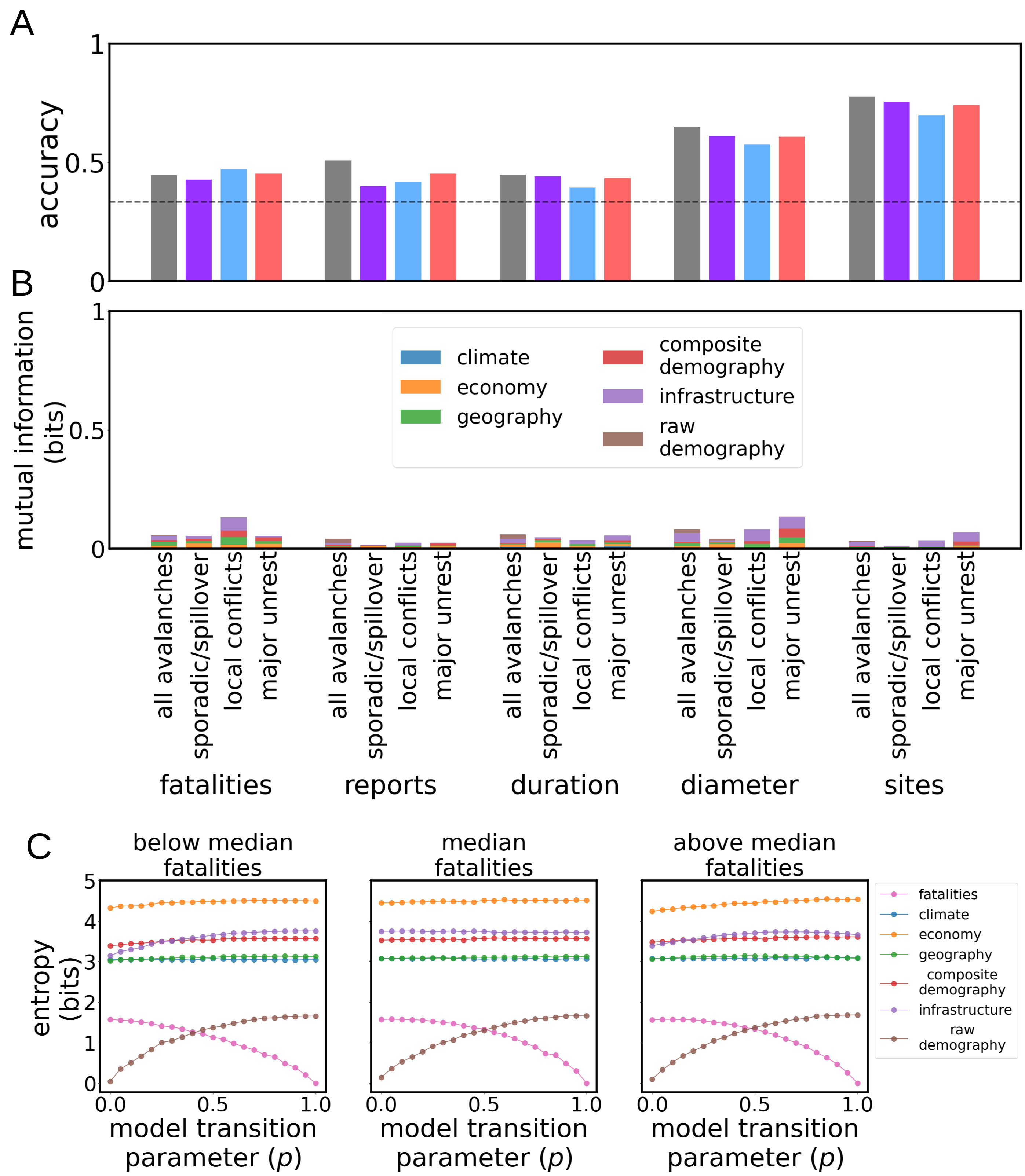}
\caption{Conflict size prediction. A) Averaged accuracy of random forest classifiers in predicting conflict avalanche size in terms of fatalities, number of reports, duration, diameter, and number of sites as below, at, or above median values \cite{leeScalingTheoryArmedconflict2020}. Bars compare model performance when trained on all conflict avalanches or on each three conflict archetype separately. Horizontal dashed line represents accuracy expected from a na\"ive random classifier. B) Mutual information between variable categories and measures of conflict size. C) Entropy trade-off between prediction of fatalities based on M4 $(p=0)$ or perfect-size predictor model ($p=1$). Pink and brown curves most clearly indicate trade-off between raw demographic value and predicting uncertainty in fatalities. See \textit{Supplementary Information} for trade-off for other measures of conflict size.}\label{fig:prediction_fig}
\end{figure}

The strength of the tripartite categorization might make us optimistic that conflict archetype could help inform useful predictions of conflict properties such as its intensity. Intensity is often measured in the number of fatalities, but analogous quantities include the number of reported events, conflict duration, diameter, and area covered---these are all measures of conflict size that would be especially useful to know with partial information of a conflict. We compute the mutual information, in Figure~\ref{fig:prediction_fig}B, between the conflict type $x$ and the several measures of conflict intensity $y$, but we find that it is very small, $I<0.1$\,bits, or that knowing the conflict type conveys little information about intensity. As a corollary, the conditional entropy $S[Y|X] = S[Y] - I[X;Y]$ gives us the uncertainty about conflict intensity $Y$ that is left over once given the conflict type $X$, which nearly saturates the maximum possible value of 1.3\,bits in all cases. To check the generality of our results, we also consider a random forest (RF) classifier, in Figure~\ref{fig:prediction_fig}A, to predict conflict intensity given the same background indicators. This step allows us to go beyond the assumptions of independence that helped facilitate the calculation of entropies and can leverage correlations between more than two variable categories. While the RF surpasses the predictive capacity based on knowing conflict type from M4, the relative improvement in performance is little better as we show in Figure~\ref{fig:prediction_fig}A.\footnote{Additionally, if we do not consider the variable categories in "bag of words" form but consider all variables individually, we find approximately a 10\% increase across the board in accuracy but no clear change in relative performance with and without conflict types (see \textit{Supplementary Information} Figure~\ref{appendixFig:prediction}).} 
This confirms our observation that knowledge of the conflict type leads to almost no reduction in the uncertainty about conflict intensity, even with variable grouping and independent treatment of variable categories.

This lack of correlation between conflict type and measures of intensity is surprising, given the focus in the literature on using such variables to regress against conflict propensity and prediction \cite{hegreViEWSPoliticalViolence2019}. Here, a lack of correlation implies that knowledge of background indicators competes with knowledge of conflict intensity. To show this trade-off, we imagine moving between two different models: M4 and a clustering algorithm that perfectly specifies conflict intensity, either fatalities, reports, etc. The latter model represents perfect correspondence between conflict intensity and conflict type. Next, we stipulate a variable $p\in[0,1]$ that determines the probability with which any given conflict avalanche is placed into its cluster as given by M4 ($p=0$) or the intensity categorization model ($p=1$). This auxiliary variable allows us to smoothly titrate between the two extremes and allows us to ``gain'' information about conflict intensity as we move away from M4.

One possible outcome is that we gain information about raw demography and fatalities simultaneously as we move from one end to the other. Other possibilities include either unchanging or loss of information about background indicators as we gain information about conflict intensity. As we show in Figure~\ref{fig:prediction_fig}C, we generally find is a strong, competitive trade-off with respect to population. The conditional entropy $S[Y|X]$ shows that as we raise the value of $p$, we must first give up essentially all knowledge of raw demographics before we gain information about conflict intensity and vice versa.\footnote{The conditional entropy corresponds to the expected error rate from a random guess given conflict type, $\epsilon\equiv1-e^{-S_y}$, which is linearly proportional to $p$, so this corresponds to an approximately linear change in error rates. In the trivial case where each cluster is assigned one and only one size variable, then this reduces to $S_y=0$, $p=1$, and $\epsilon=0$.} Furthermore, the other variables depend weakly on $p$, and also show increase in the conditional entropy or remain flat. Thus, another way to characterize the trade-off is that information gain about raw demographics leads directly to a loss of information about conflict intensity and no information is gained from considering the other variables. The competitive trade-off is a result of independence, or that knowledge of conflict category as found here hurts (and at best does not improve) knowledge of conflict intensity across all measures of conflict intensity for all variable types.

\section*{Discussion}
Armed conflicts are often categorized into separate types using identifiable mechanisms via which they start, develop, and terminate. Such distinctions are presumably informative because they relate conflict properties within the circumscribed set that do not generalize to without \cite{midlarskyHandbookWarStudies2011}.
The implicit (and intuitive) notion is that the way that variables depend on one another are more similar between certain types of conflicts than between others. Indeed, similarities along economic, organizational \cite{weinsteinRebellionPolitics2007}, political, and historical aspects of conflict have motivated insightful theoretical frameworks and typologies. Understanding the typology of armed conflict is critical for developing political theories tailored to different conflict types, grasping the underlying mechanisms of conflict ignition, and establishing effective policy interventions \cite{angstromTypologyInternal2001}. Here, we discover such types from data.

To address the challenge holistically, we develop a vertically integrated and empirically-based procedure. We first construct chains of conflict events, ``conflict avalanches,'' from disaggregated conflict data across a wide range of scales in spatiotemporal resolution as in Figure~\ref{fig:avalanches} \cite{raleighIntroducingACLEDArmed2010,kushwahaDiscoveringMesoscaleChains2023a}. 
Then, we aggregate highly-resolved data sets on conflict drivers or proxies thereof to measure conflict avalanche attributes (Figure~\ref{fig:data}) that feed into an unsupervised learning technique, a variation on the multinomial mixture model. We identify three major types of conflict, which collectively constitute a ``triangle of madness'' (Figure~\ref{fig:proj}). 

The three conflict types represent overarching classes, or archetypes, which we name ``major unrest,'' ``local conflicts,'' and ``sporadic and spillover events'' based on typical avalanche properties. 
When we take well-documented examples of conflict, we find that conflicts of similar intensities appear in the same cluster. For example, the Al-Shabaab insurgency \cite{andersonUnderstandingAlShabaabClan2015}, Boko Haram insurgency \cite{walkerWhatBokoHaram2012} and the Central African Republic civil war \cite{arieff2014crisis} all fall within major unrest. Similarly, conflicts in Ituri \cite{faheyTroubleIturi2011} and Kivu \cite{ekyambaAssessingChallengesArmed2022}, the Seleka and anti-Balaka conflict \cite{kah2014anti,isaacs-martinSelekaAntiBalakaRebel2017}, and local clan violence in Somalia (2003–2004) are in the local conflicts. Additionally, extensions of the Al-Shabaab insurgency in Somalia, such as the conflict around Bosaso \cite{SomaliaAlShabaabAttack2017}, are categorized under sporadic and spillover. 
That these examples cluster together as political analysis would suggest validates the data-oriented and unsupervised approach, drawing on statistical similarities in population, infrastructure, economic conditions, etc.

The alignment in background conditions, however, does not necessarily imply that the conflict types are separated geographically. Inspecting the geographic distribution of conflicts (Figure \ref{fig:proj}E), we find that the types often touch or overlap. For instance, the area surrounding Mogadishu exhibits major unrest associated with the longstanding Al-Shabaab insurgency alongside local conflicts of brief duration and numerous sporadic and spillover events. Another interesting area is the tri-border region of Burundi, Rwanda and the Democratic Republic of the Congo. We see that the major unrest cluster contains a large-scale conflict avalanche that extends across national boundaries and persists over an extended duration, shown in color green in Figure~\ref{fig:proj}E. In contrast, within the local conflicts cluster, the same region exhibits smaller-scale conflict avalanches that are confined within individual national borders, also shown in Figure~\ref{fig:proj}E. Furthermore, long conflicts part of the instability in the Maghreb region \cite{planetcontrerasRecentHistory2007} are all part of major unrest cluster and small conflicts of Maghreb are part of the local conflict cluster. These examples underscore the variability in conflict dynamics within the same or adjacent regions, where certain events escalate into major unrests, while others remain local or sporadic.

The ``triangle of madness'' hierarchically splits into a taxonomy of conflict types (Figure~\ref{fig:hierarchy}). The empirical taxonomy indicates a particular order in which four of the variable categories seem to be most important for identifying conflict types: raw demographics, infrastructure, economics, and lastly geography. Climate plays a smaller role compared to other factors, such as economics, consistent with other studies \cite{machClimateRiskFactor2019, ayanaExaminingRelationshipEnvironmental2016,wischnathClimateVariabilityCivil2014}.
Unlike heuristic classifications of armed conflicts, which often take the form of conceptual typologies \cite{baileyTypologiesTaxonomiesIntroduction2003}, our approach is grounded in empirical data. 

Despite these clear division into the three archetypes, we find little information between archetype and measures of conflict intensity (fatalities, reports, duration, spatial spread) under a predictive test. Indeed, a general measure of nonlinear dependency, the mutual information, is small (Figure~\ref{fig:prediction_fig}B).
This implies that such specification is detrimental to predicting conflict intensity. This competitive effect is a general consequence of weak dependence and holds across scales of conflict analysis and other ways of grouping conflict events including by country (see \textit{Supplementary Information} Figure~\ref{appendixFig:MI_country}). We confirm our findings with the random forest model to go beyond assumptions that we made to estimate the mutual information. Although our model's overall accuracy surpasses that of a random classifier, its performance generally declines—or, at best, exhibits only marginal improvement—when conflict archetypes are evaluated individually (Figure \ref{fig:prediction_fig}A). This implies that the observation is not specific to our procedure, but likely represents an important cost to using certain background indicators.

In this sense, our work touches on ongoing work on armed conflict prediction. Our findings indicate that while commonly available background indicators present strong patterns (which may even help frame policy), such clarity does not necessarily predict conflict properties. 
This is indicative of the wider challenge of quantitative conflict prediction \cite{cedermanPredictingArmedConflict2017} in which strong prediction has been elusive. For example, a particularly visible and comprehensive approach is based on a dynamic multinomial logit model \cite{hegrePredictingArmedConflict2013, obukhovIdentifyingConditioningFactors2023}, but such techniques fall well short of true positive rates of 50\% for the incidence of conflict with overall performance mostly dominated by conflict infrequency \cite{hegreCanWePredict2021}. In alignment with our findings, incorporating geography slightly improves the prediction accuracy by reducing the false positive rates \cite{weidmannPredictingConflictSpace2010}. Even when considering a wider pool of algorithms---such as from a prediction competition hosted by the Violence \& Impacts Early-Warning System (VIEWS) \cite{hegreLessonsEscalationPrediction2022}---the general observation is that predictions are limited in accuracy and precision \cite{cedermanPredictingArmedConflict2017, friedmanWarChanceAssessing2019}. 
This leads directly to the question of the utility of data sources. In at least one example, highly context-specific, open-source information seems to have been successfully deployed in Afghanistan \cite{spahrRavenSentry2024}---although this success was also predicated on human intelligence and lack of published details make its generality hard to assess. Newer techniques like the text-based actor embeddings to predict conflict dynamics may go beyond predictive models using solely background indicators \cite{croicuNewswireNexus2025}. Looking ahead, prediction may ultimately depend much more on expanding on the set of background indicators than about squeezing the (minuscule) signal available in common ones.

\acknow{We thank Jan Fialkowski for helpful discussion. EDL acknowledges funding from the Austrian Science Fund grant ESP-127. NK acknowledges funding from the Austrian Federal Ministry for Climate Action, Environment, Energy, Mobility, Innovation and Technology (2021-0.664.668) and the City of Vienna.}

\showacknow{} 


\bibliography{clustering_armed_conflict}

\clearpage
\appendix

\renewcommand{\thefigure}{A\arabic{figure}}  
\setcounter{figure}{0}  

\renewcommand{\thesubsection}{\thesection.\arabic{subsection}}

\titleformat{\section}
  {\normalfont\Large\bfseries} 
  {Appendix \thesection} 
  {0.5em} 
  {#1}

\onecolumn

\begin{center}
    \Huge\bfseries Supplementary Information
\end{center}

\tableofcontents
\clearpage
\twocolumn

\section{Heuristic classifications}\label{appendix:heuristic classification}
The following classifications of armed conflicts was originally compiled in \cite{breckeAidFinding1997}.
\subsection*{Inter-state conflict}
\begin{itemize}
    \item international war
    \item global war
    \item world war
    \item general war
    \item systemic war
    \item major coalition war
    \item major powers war
    \item war of rivalry
    \begin{itemize}
        \item hegemonic war
        \item power transition war
        \item status war
        \item colonial war (between colonial occupiers)
    \end{itemize}
    \item territorial conflict
    \begin{itemize}
        \item border war (between countries)
        \item border skirmish
        \item navigation war
        \item territorial dispute
        \item frontier conflict
    \end{itemize}
    \item state-sponsored terrorism (in other countries)
    \item subversion
    \item irredentist conflict
    \item counter-revolutionary war
    \item armed attack
    \begin{itemize}
        \item invasion
        \item missile attack
        \item bombing attack
        \item bombing campaign
        \item bombardment
    \end{itemize}
    \item intervention
    \item occupation of territory
    \item expansionist war
    \item collaborationist conflict
    \item neo-colonial conflict
\end{itemize}

\subsection*{Conflict between state and external non-state actor}
\begin{itemize}
    \item extra-systemic war
    \item imperial war
    \item colonial war
    \item war of liberation
    \item war of independence
    \item revolutionary war
    \item decolonization conflict
    \item armed rebellion
    \item colonial liberation war
    \item state-building war (expanding into ``unoccupied'' territory)
    \item colonial expansion war
    \item war of resistance
    \item war of occupation
    \item drug war
\end{itemize}

\subsection*{Intra-state conflict}
\begin{itemize}
    \item civil war
    \item revolution
    \begin{itemize}
        \item political revolution
        \item social revolution
        \item urban revolution
        \item peasant revolution
        \item palace revolution
        \item millenarian revolution
        \item anarchistic revolution
    \end{itemize}
    \item state-building war
    \item state formation conflict
    \item insurgency
    \begin{itemize}
        \item armed insurgency
    \end{itemize}
    \item rebellion
    \begin{itemize}
        \item armed rebellion
    \end{itemize}
    \item revolt
    \begin{itemize}
        \item peasant revolt
        \item armed revolt
    \end{itemize}
    \item peasant war
    \item peasant rebellion
    \item jacquerie
    \item coup d'etat/putsch
    \begin{itemize}
        \item palace coup
        \item reform coup
        \item revolutionary coup
        \item conspiratorial coup d'etat
    \end{itemize}
    \item purge
    \item pronunciamento
    \item dynastic war
    \item war of succession
    \item terrorism
    \begin{itemize}
        \item attacks to cripple economy
        \item attacks to shake faith in government
    \end{itemize}
    \item ethnic conflict
    \begin{itemize}
        \item ethno-political conflict
        \item race conflict
        \item race war
    \end{itemize}
    \item expulsion
    \item group identity conflict
    \item war of self-determination
    \item war of secession
    \item insurrection
    \begin{itemize}
        \item secessionist armed insurrection
        \item armed insurrection
        \item militarized mass insurrection
    \end{itemize}
    \item uprising
    \begin{itemize}
        \item armed uprising
        \item peasant uprising
    \end{itemize}
    \item conflict to achieve limited self-rule
    \item separatism
    \item genocide
    \item politicide
    \item massacre
    \item government repression of social groups
    \item state terrorism
    \item government oppression
    \item pogrom
    \item counter-terrorism campaign
    \item warlord battles for control of collapsed state
    \item clan warfare
    \item factional warfare
    \item internecine warfare
    \item class conflict
    \begin{itemize}
        \item class warfare
    \end{itemize}
    \item state resistance conflict
    \item riot
    \item land seizure
\end{itemize}

\subsection*{Abstract properties}
\begin{itemize}
    \item simple conflict
    \item complex conflict
    \item recurring conflict
    \item low intensity conflict
    \item guerilla war
    \item trench warfare
    \item weapons of mass destruction war
    \item proxy war
    \item local war
    \item regional war
    \begin{itemize}
        \item regional internal war
    \end{itemize}
    \item relative deprivation conflict
    \item cultural conflict
    \item distributive dispute
    \item ideological conflict
    \item personnel war
    \item authority war
    \item structural war
\end{itemize}

\subsection*{Either inter-state or intra-state}
\begin{itemize}
    \item ideological war
    \item political war
    \item post-colonial war
    \item religious conflict
    \begin{itemize}
        \item religious war
    \end{itemize}
    \item environmental conflict
    \begin{itemize}
        \item scarcity conflict
        \item resource conflict
        \item pollution/emissions conflict
    \end{itemize}
\end{itemize}

\subsection*{Borderline violent conflict}
\begin{itemize}
    \item incident
    \item clash
    \begin{itemize}
        \item armed clash
    \end{itemize}
    \item agitation
    \item unrest
    \item disturbance
    \item disorder
    \item mutiny
    \item piracy
\end{itemize}

\onecolumn 

\section{Datasets}\label{appendix:data}
This section contains information about all the datasets that we use in this paper. Information about armed conflict events comes from the Armed Conflict Location and Event Database (ACLED). Apart from this, we have datasets spanning six factors that are often associated with armed conflicts. These factors are climate, economy, infrastructure, geography, composite demography and raw demography. The datasets were selected such that they satisfy certain constraints needed for this analysis. These constraints were:
\begin{itemize}
    \item The dataset should be updated temporally atleast annually (except geography which can be static)
    \item The dataset should be in raster format with high spatial resolution.
    \item All the datasets should be available for a common time period which also coincides with the ACLED data available to us.
    \item should be publicly available for free.
\end{itemize}
Observing these constraints, we were able to collect 22 datasets between the years 2000-2015. Here are the summary of those datasets.


\subsection{ACLED}\label{appendixSub:ACLED}
Our primary dataset is the Armed Conflict Location \& Event Data (ACLED) Project. This project collects data on armed conflicts around the world with a focus on African states. The dataset is a collection of individual conflict events, defined as a single incidence of violence at a particular location and time involving at least two actors. In our analysis, we primarily focus on the location and date of the conflict events, and we use other information including actor identities and event description for validation of the conflict avalanches.

Other event-based armed conflict datasets besides ACLED include the Global Terrorism Database (GTD); the Integrated Crisis Early Warning System (ICEWS) dataset; the Phoenix event dataset; the Global Database of Events, Language, and Tone (GDELT); and the Uppsala Conflict Data Programme Georeferenced Event Dataset (UCDP GED) \cite{raleigh2019comparing}.
We choose to use ACLED in our analysis because of two major reasons:
\begin{enumerate}
    \item Event-based armed conflict databases extract their information from various news reports from multiple sources. This can be done either manually by the help of human researchers and experts or can be scraped automatically from news articles. Since we are focusing on Africa, we require a dataset which is curated manually by experts since most news articles published in Africa are not in English and should have some understanding of local context. ACLED, GTD, and UCDP GED are the only three expert-curated datasets. The others are compiled using automated systems which tend to be heavily biased towards conflict events reported in English and French media \cite{raleigh2019comparing} since currently automated systems are not designed to crawl through local language media.
    \item ACLED covers all violent activities that occur both within and outside the context of a civil war, particularly violence against civilians, militia interactions, communal conflict, and rioting. The other data sets do not. GTD focuses on ``terrorism'' only. 
    UCDP GED only records conflict events with at least one fatality. These definitions of armed conflicts are too restrictive for our purposes. Therefore, ACLED is the most suitable dataset for our analysis among the available event-based datasets.
\end{enumerate}

\subsection{Temperature}

We utilize temperature data from the Climatic Research Unit Gridded Time Series (CRU TS) dataset (\cite{harrisVersionCRUTS2020}). This dataset comprehensively records 2m temperature and various other meteorological variables of land surfaces, excluding Antarctica. The data is derived from weather station observations, that undergo a homogenization process to ensure accuracy and consistency. Each grid within the dataset provides daily time series data of mean temperature from 1901 onward and is represented at a spatial resolution of $0.5^{\circ}\times0.5^{\circ}$ grid cells.

Click \href{https://crudata.uea.ac.uk/cru/data/hrg/cru_ts_4.07/cruts.2304141047.v4.07/tmp/cru_ts4.07.1901.2022.tmp.dat.nc.gz}{here} to access the data.

\subsection{Precipitation}

Our precipitation dataset is sourced from the Climate Hazards Group InfraRed Precipitation with Station (CHIRPS) dataset \cite{funkClimateHazardsInfrared2015a}. This dataset combines satellite-based infrared observations with ground-based station data to produce precise precipitation estimates. Notably, it excels in providing fine-scale spatial resolution at $0.05^{\circ}$. The dataset covers from 1981 and offers various temporal scales, including 6-hourly, daily, pentad, dekad, and monthly intervals. It is widely recognized for its strength in drought monitoring \cite{pengPanAfricanHighresolutionDrought2020,dasEvaluatingAccuracyTwo2022,gebrechorkosAnalysisClimateVariability2020}.

Click \href{https://data.chc.ucsb.edu/products/CHIRPS-2.0/global_daily/netcdf/p25/}{here} to access the data.

\subsection{Vegetation}

Normalized Difference Vegetation Index (NDVI) is an indicator that quantifies the density and health of vegetation through remote sensing. This indicator ranges from -1 to 1, with values close to 1 indicating high greenness due to dense or healthy vegetation conditions (e.g., tropical forests, cropland). As NDVI values approach 0, the land becomes more barren (e.g., desert). NDVI values of deep water bodies are detected as -1. In this study, we used daily NDVI data from the National Oceanic and Atmospheric Administration (NOAA) at a spatial resolution of $0.05^{\circ}$ (available from 1981 to the present). 

Click \href{https://www.ncei.noaa.gov/thredds/catalog/cdr/ndvi/catalog.html}{here} to access the data.

\subsection{GDP, GDP per capita and HDI}

Kummu et al. \citep{kummuGriddedGlobalDatasets2018} published a gridded database of economic and human development indicators, such as GDP, GDP per capita, and HDI, at both national and sub-national levels. Missing values were filled through interpolation and extrapolation. For this study, we used the African data from 1990 to 2015 at a 5 arc-min spatial resolution.

Click \href{https://datadryad.org/stash/dataset/doi:10.5061/dryad.dk1j0}{here} to access the data.

\subsection{Distance from inland water bodies}

WorldPop and Lamarche et al. \citep{lamarcheCompilationValidationSAR2017} provide a dataset that calculates the closest geodesic distances between grid cell centers and inland waterbodies at the 3-arc second resolution. This dataset does not capture changes over time, but the distance values are for the 2000-2012 period.

Click \href{https://hub.worldpop.org/geodata/listing?id=61}{here} to access the data.

\subsection{Distance from coastline}

WorldPop offers an open-access dataset that includes the closest distances between the open-water coastline and the center of grid cells at the 3-arc second resolution. The temporal scale of the data is invariant for the 2000-2020 period.

Click \href{https://hub.worldpop.org/geodata/listing?id=60}{here} to access the data.

\subsection{Elevation}

The elevation data in 2000 (above the sea level) is accessible through WorldPop at a 3-arc second resolution. This dataset is derieved from the NASA's Shuttle Radar Topography Mission (SRTM) data.

Click \href{https://hub.worldpop.org/geodata/listing?id=58}{here} to access the data.

\subsection{Net migration, Birth rate and Death rate}

Niva et al. \citep{nivaWorldHumanMigration2023} used STATcompiler (only birth), Eurostat, OECD regional stats (only death), and census data to estimate subnational birth and death rates annually for 1990-2019. These rates are calculated as the number of births or deaths per 1000 populations. The calculation includes the WorldPop population data and HYDE 3.2 data (see A.15 for more details on the population data). Birth and death rates are downscaled based on the HDI, population density, the ratio of women of the reproductive age, and the proportion between average age and life expectancy. Natural population change is calculated as deaths minus births, while net migration is obtained by subtracting natural population change from total population change.

Click \href{https://zenodo.org/record/7997134}{here} to access the data.

\subsection{Interacting ethnic groups}

Vogt et al. \citep{vogtIntegratingDataEthnicity2015} published the Ethnic Power Relations (EPR) dataset, the first version in 2014 and the updated version in 2021. Within the EPR dataset, the GeoEPR dataset provides geospatial information on ethnic groups. The dataset can also track how the geographical bases of ethnic groups change over time.

Click \href{https://icr.ethz.ch/data/epr/geoepr/}{here} to access the data.

\subsection{Cellular phone per 100 people}

The World Bank collects mobile cellular telephone subscription data per 100 people for public mobile telephone service. The data includes the number of postpaid subscriptions and active prepaid accounts in the past three months while missing certain types of subscriptions (e.g., subscriptions by data cards/USB modems, subscriptions to public mobile data services, etc.). The data are based on the administrative data of telecommunication authorities, government offices, or operators---vary by country. Note that the data quality differs across countries depending on the local situations, e.g., data availability, telecommunication regulation.

Click \href{https://data.worldbank.org/indicator/IT.CEL.SETS.P2}{here} to access the data.

\subsection{Electric consumption}

Chen et al. \citep{chen2022global} presents a global gridded dataset of electricity consumption at a 1 km x 1 km resolution from 1992 to 2019. The dataset is constructed based on the calibrated nighttime light data. This dataset overcomes the limitations of existing electricity consumption data, capturing realistic GDP growth, varying spatiotemporal dynamics, and restricted temporal coverages.

Click \href{https://figshare.com/articles/dataset/Global_1_km_1_km_gridded_revised_real_gross_domestic_product_and_electricity_consumption_during_1992-2019_based_on_calibrated_nighttime_light_data/17004523/1}{here} to access the data.

\subsection{Shortest distance to a road}
We calculated the shortest distance from the center of each Voronoi cell to the Global Roads Open Access Data Set Version 1 (gROADSv1) provided by the Center for International Earth Science Information Network (CIESIN), Columbia University. gROADSv1 data has varying temporal and spatial scales due to its development from multiple sources.

Click \href{https://www.earthdata.nasa.gov/data/catalog/sedac-ciesin-sedac-groads-v1-1.00}{here} to access the data.

\subsection{Night light}

Li et al. \citep{liHarmonizedGlobalNighttime2020} integrated two sets of night light data collections from two sensors with different temporal coverages, VIIRS-DNB for 2012-2020 and DMSP-OLS for 1992-2013. The spatial resolution is 30 arc seconds, and the temporal resolution is daily from 1992 to 2020.

Click \href{https://figshare.com/articles/dataset/Harmonization_of_DMSP_and_VIIRS_nighttime_light_data_from_1992-2018_at_the_global_scale/9828827/7}{here} to access the data.

\subsection{Population}

WorldPop offers yearly population counts and density data at the spatial resolution of 30 arc. The data was mapped by the Random Forest-based dasymetric redistribution.

Click \href{https://hub.worldpop.org/geodata/listing?id=75}{here} to access the data (count).

Click \href{https://hub.worldpop.org/geodata/listing?id=77}{here} to access the data (density).

\section{M4 model}\label{appendix:m4}
\subsection{Notation}
Note: The notation used in the following derivation is different from the one used in the main paper.\\
\begin{itemize}
    \item Data: $i=1,2,...,M$
    \item Clusters: $j=1,2,...,K$
    \item Components/variable categories: $c=1,2...,S$
    \item Number of possible outcomes/number of divisions used in coarse-graining conflict avalanche vector: $L$
    \item $Mult(\vec{\theta},N)$ is a multinomial distribution with parameters $\vec{\theta}$ and N total draws such that $n_1+n_2+... = N$
\end{itemize}

\subsection{Multinomial mixture model}
A mixture model is a probabilistic model to soft cluster data where each data point is said to be sampled from a mixture of some distributions. If the base distribution used is a multinomial distribution, the model is called a multinomial mixture model (M3), 

\begin{align*}
        P(x_i|\vec{\theta}) &= \sum_{j}^{K} \pi_j Mult(\vec{\theta}_{j}, N_{i}) \\
        &= \sum_{j}^{K} \pi_j \left(\frac{N!}{\prod_{a}^L n_{i,a}!} \prod_{a=1}^L \theta_a^{n_{i,a}} \right)
\end{align*}

Multinomial mixture models are predominantly utilized for classifying documents into distinct topics \cite{novovicovaApplicationMultinomialMixture2003}, where each topic is characterized by a distribution over unique words (for example, the topic \textit{sports} will assign a higher weight to the word "basketball" compared to the topic \textit{food}). In this modeling framework, the words within a document are considered independent samples drawn from a topic distribution. As an unsupervised clustering method, the multinomial mixture model requires pre-specification of the desired number of clusters (or topics in the context of document clustering). Upon setting this hyper-parameter , the model clusters the data into the specified number of clusters. Applying this analogy to our analysis, each conflict avalanche is treated as a \textit{document}. The conflict avalanches are then grouped into different \textit{topics} or clusters using a variation of the standard multinomial mixture model which we call the multi-multinomial mixture model (M4).

In this new variation of M3, each multinomial is replaced by the product of multinomials where each multinomial distribution is associated with each variable category (see next section for equation).
We employ the M4 to fit our conflict avalanche vectors, determining the optimal fit by evaluating the maximum log likelihood across 1000 model fits. The M4 provides probabilities indicating the likelihood of a particular conflict avalanche belonging to a specific cluster, thereby facilitating a soft clustering. To derive a hard clustering, we assign a cluster label to each conflict for which the probability surpasses the 0.5 threshold.

According to M4, the probability that a given conflict avalanche $\vec{x_i}$ belongs to cluster $j$ is given by,
\begin{align}
    P(\vec{x_i},z_i=j|\vec{\theta}) = \pi_j\prod_{c=1}^{s} Mult(\vec{\theta}_{j}^{c}, N_{}^{c})
\end{align}
Here, $z_i$ is the cluster indicator for conflict avalanche $\vec{x_i}$ and $\pi_j$ is the probability of selecting cluster $j$ out of the $k$ clusters, which is proportional to the total number of conflict avalanches assigned to cluster $j$. $Mult(\vec{\theta}_{j}^{c}, N_{}^{c})$ denotes a multinomial distribution parameterized by $\vec{\theta_{j}^{c}}$ and total number of variables $N_{}^{c}$ of variable type $c$. $s$ equals the total number of variable categories, six in our case. $\vec{\theta_{j}^{c}} =  \left \{ \theta_{j,\uparrow}^{c}, \theta_{j,\approx}^{c}, \theta_{j,\downarrow}^{c} \right \}$ is the probability of sampling a value either below median, at median or above median range for a variable of variable type $c$ for cluster $j$. $n_{i,\uparrow}^{c} + n_{i,\approx}^{c} + n_{i,\downarrow}^{c} = N^c$ where $n_{i,\uparrow}^{c}$, $n_{i,\approx}^{c}$, $n_{i,\downarrow}^{c}$ are the number of variables of variable type $c$ which have above median, at median and below median values respectively for conflict avalanche $\vec{x_i}$.
Drawing an analogy to document classification, this equation suggests that each variable category is analogous to a \textit{sub-topic} and the product of these six \textit{sub-topic} distributions gives us the \textit{topic} or cluster distribution\footnote{By setting the cluster distribution to be a product of six distinct variable type distributions, our model implicitly adopts the assumption of independence among these variable types at the start of model fitting.}.

\subsection{M4 fitting using expectation-maximization(EM)}

The M4 is given by,

\begin{align}
        P(x_i|\vec{\theta}) = \sum_{j}^{K} r(z_i=j) P(x_i|z_i=j,\vec{\theta})  
\end{align}

where $z_i$ represents the latent class/cluster for data point $x_i$ and,
\begin{align*}
        r(z_i)= Mult(\vec{\pi},1) \\
        r(z_i=j) = \pi_j \\ 
        P(x_i|z_i=j,\vec{\theta}) = \prod_{c}^{s} Mult(\vec{\theta}_{j}^{c}, N_{i}^{c}) \\
        \vec{\theta}_{j}^{c} =  \left\{\theta_{j,1}^c, \theta_{j,2}^c,...,\theta_{j,L}^c\right\}
\end{align*}

such that,
\begin{align*}
    P(x_i,z_i=j|\vec{\theta}) = \pi_j\prod_{c}^{S} Mult(\vec{\theta}_{j}^{c}, N_{i}^{c})
\end{align*}

Let,
\begin{align*}
    P(z_i=j|x_i,\vec{\theta})=\tau_{ij}
\end{align*}

The log likelihood estimator will be,
\begin{align*}
    Q &=\sum_i^M\sum_j^K P(z_i=j|x_i,\vec{\theta}) \log{P(x_i,z_i|\vec{\theta})} \\
    &= \sum_i^M\sum_j^K \tau_{ij} \log{\left (  \pi_j \prod_{c}^{S} Mult(\vec{\theta}_{j}^{c}, N_{i}^{c})\right )} \\
    &= \sum_i^M\sum_j^K \tau_{ij} \left \{ \log{\pi_j} + \sum_c^S{ \log{Mult(\vec{\theta}_{j}^{c}, N_{i}^{c})}} \right \}
\end{align*}

Incorporating the constraints to apply Lagrange's multiplier method we get,
\begin{align*}
    Q' &= \sum_i^M\sum_j^K \tau_{ij} \left \{ \log{\pi_j} + \sum_c^S{ \log{Mult(\vec{\theta}_{j}^{c}, N_{i}^{c})}} \right \} - \lambda_\theta\left \{ 1-\sum_a^d \theta_{ja}^c \right \}-\lambda_\pi\left \{ 1-\sum_j^K \pi_j \right \} \\
    &= \sum_i^M\sum_j^K \tau_{ij} \left \{ \log{\pi_j} + \sum_c^S \left \{ log{\frac{N_i^c!}{n_{i,1}^c!...n_{i,L}^c!} + \sum_a^L \log\left ( \theta_{ja}^c \right )^{n_{ia}^{c}}} \right \}  \right \}     - \lambda_\theta\left \{ 1-\sum_a^L \theta_{ja}^c \right \} -\lambda_\pi\left \{ 1-\sum_j^K \pi_j \right \}
\end{align*}

Therefore,
\begin{align}
\label{lagrange1}
    \frac{\partial Q'}{\partial \theta_{ja}^c} = \sum_i^M \tau{ij}\left \{ \frac{n_{ia}^c}{\theta_{ja}^c} \right \} + \lambda_\theta \\
    \label{lagrange2}
    \frac{\partial Q'}{\partial \pi_j} = \sum_i^M \tau{ij}\left \{ \frac{1}{\pi_j} \right \} + \lambda_\pi
\end{align}

Using \ref{lagrange1},
\begin{align}
    \label{theta1}
    \sum_i^M \tau_{ij} n_{ia}^c = -\lambda_\theta \theta_{ja}^c \\
    \label{theta2}
    \Rightarrow \sum_i^M \sum_a^L \tau_{ij} n_{ia}^c = -\lambda_\theta 
\end{align}

Substituting \ref{theta2} into \ref{theta1},
\begin{align}
    \theta_{ja}^c = \frac{\sum\limits_{i}^{M} \tau_{ij} n_{ia}^c}{\sum\limits_{i}^{M} \sum\limits_{a}^{L} \tau_{ij}n_{ia}^c} \\
    \label{theta}
    \Rightarrow \theta_{ja}^c = \frac{\sum\limits_{i}^{M} \tau_{ij}n_{ia}^c}{\sum\limits_{i}^{M} \tau_{ij}N_i^c} 
\end{align}

Using \ref{lagrange2},
\begin{align}
    \label{pi1}
    \sum_i^M \tau_{ij} = - \lambda_\pi \pi_j \\
    \label{pi2}
    \Rightarrow \sum_i^M \sum_j^K \tau_{ij} = - \lambda_\pi
\end{align}

Substituting \ref{pi2} into \ref{pi1},
\begin{align}
    \pi_j = \frac{\sum\limits_{i}^{M} \tau_{ij}}{\sum\limits_i^M \sum\limits_j^K \tau_{ij}} \\
    \label{pi}
    \Rightarrow \pi_j = \frac{\sum\limits_{i}^{M} \tau_{ij}}{M}
\end{align}

\ref{theta} and \ref{pi} is the update rule from the M-step of EM algorithm.

The update rule from the E-step is,
\begin{align}
    \tau_{ij} = P(z_i=j|x_i,\vec{\theta}) = \frac{P(x_i,z_i=j|\vec{\theta})}{P(x_i|\vec{\theta})} \\ 
    \label{tau}
    \Rightarrow \tau_{ij} = \frac{\pi_j \prod\limits_c^S Mult(\vec{\theta}_j^c,N_i^c)}{\sum\limits_j^K \pi_j \prod\limits_c^S Mult(\vec{\theta}_j^c,N_i^c)} 
\end{align}

\section{Generating vectors of conflicts}\label{appendix:conflict_vector}
\begin{figure}[h]\centering
    \includegraphics[width=\linewidth, height=\textheight, keepaspectratio]{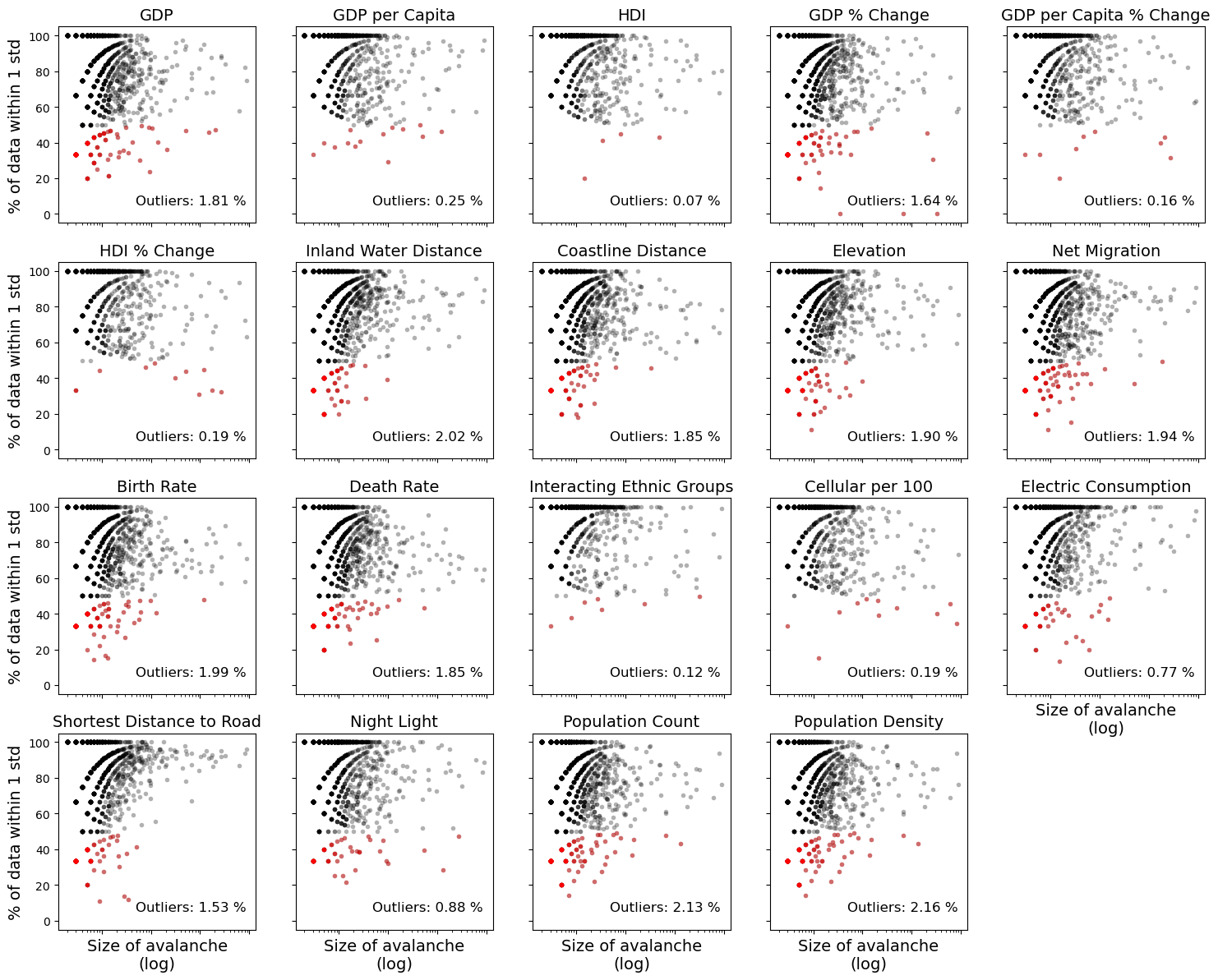}
    \caption{This analysis evaluates whether conflict avalanches can be effectively characterized by the mean values of a given variable computed across the events within each avalanche. Each point in the figure represents a single conflict avalanche. The x-axis displays the percentage of events within an avalanche whose value for the variable falls within one standard deviation of the overall distribution of that variable in the avalanche. Avalanches in which fewer than 50 percent of events lie within one standard deviation are marked in red and are called outliers. Variations in darkness of point colors indicate overlapping points. The results suggest that, for the vast majority of conflict avalanches, the values of the variable are highly consistent (i.e., within one standard deviation), thereby justifying the use of the average value to represent the entire avalanche.}\label{appendixFig:within_std}
        \addcontentsline{toc}{subsection}{Figure \ref{appendixFig:within_std}: Variance within conflict events in conflict avalanches}
\end{figure}

\begin{figure*}[h]\centering
    \includegraphics[width=\linewidth, height=\textheight, keepaspectratio]{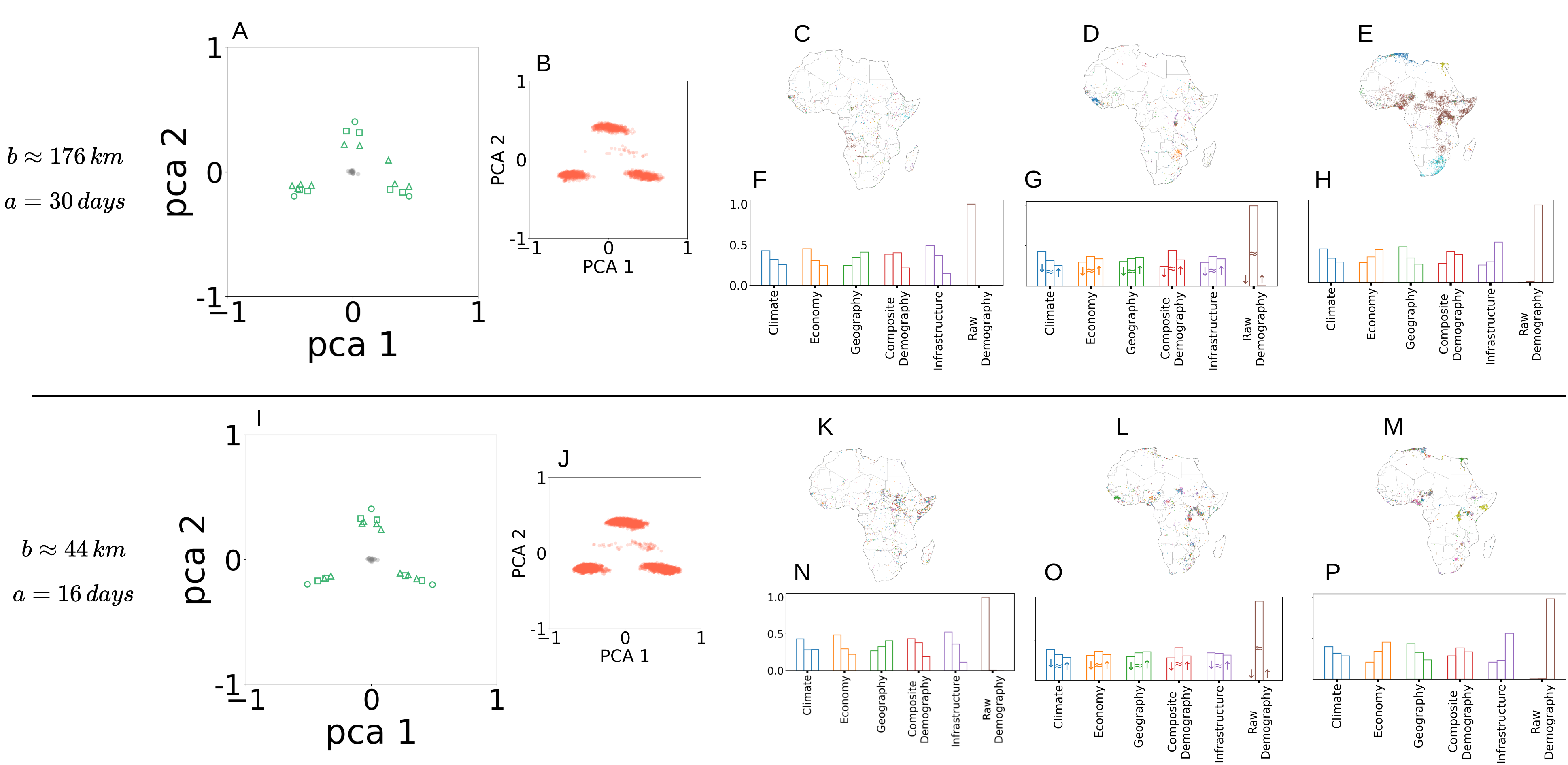}
    \caption{The triangle of madness exists at multiple scales. Here we show two representative scales which are $b=176\,km,a=30\,days$ and $b=44\,km,a=16\,days$. PCA biplots A) and I) show that three conflict archetypes emerge at these scales too, similar to one seen in Figure~\ref{fig:proj} of the main text. B) and J) shows avalanches projected to the same PCA space as panel A and panel I. C),D) and E) shows the avalanches in each cluster at $K=3$ for $b=176\,km,a=30\,days$ scale along with their model parameters in F),G) and H). K), L) and M) shows the avalanches in each cluster at $K=3$ for $b=44\,km,a=16\,days$ scale along with their model parameters in N),O) and P). The bar plots show $\vec{\theta_j^c}$ corresponding to each cluster, with three bars corresponding to each variable category. These bars denote the tendency of variables within each variable category, for that cluster, to fall below, at, or above the median, highlighted by $\downarrow, \approx$ and $\uparrow$ respectively}\label{appendixFig:PCA_SI_2}
\addcontentsline{toc}{subsection}{Figure \ref{appendixFig:PCA_SI_2}: Triangle of madness at different scales}
\end{figure*}

Below, we detail the methodology used to construct discrete conflict avalanche vectors. (Readers interested solely in the algorithm may refer to Section~\ref{appendix:vector_algorithm}.).
\\
\begin{itemize}
\item \textbf{Data gathering}\\
\\
In this project, we employed a diverse set of datasets. A comprehensive overview of all the datasets utilized is provided in Appendix~\ref{appendix:data}. In addition to the ACLED dataset, which serves as our primary source for armed conflict events, we incorporated several datasets corresponding to factors frequently associated with armed conflicts in the literature. These datasets are categorized into six groups: climate, geography, composite demography, infrastructure, economy, and raw demography. All datasets cover the time period from 2000 to 2015, which thereby defines the temporal scope of this study.
\\
\item \textbf{Conflict event and data mapping}\\
\\
Each conflict event in the ACLED dataset was mapped to the corresponding data points from the other datasets. The ACLED dataset provided precise geographic coordinates (latitude and longitude) and the exact date for each conflict event, enabling the spatial and temporal alignment of auxiliary data. For the majority of the datasets, which were rasterized, the mapping was straightforward. Raster data were associated with conflict events using the lowest available temporal resolution of each dataset. For instance, if a dataset was available on an annual basis, all conflict events occurring at the same geographic coordinates within the same year were assigned the same value; similarly, if a dataset was available monthly, events occurring in the same month and location received the same value. In certain cases, additional processing was necessary. For example, the shortest distance to roads was computed manually using a road network dataset, and the total number of ethnic groups at each conflict location was determined from the GeoEPR dataset \cite{vogtIntegratingDataEthnicity2015}. Following meticulous mapping and verification of each dataset, we obtained a unified dataframe in which each row corresponds to a conflict event and each column contains associated information (e.g., GDP, shortest distance to roads, NDVI, elevation, population count, etc.; see Figure~\ref{fig:data} in the main text for a complete list of variables).
\\
\item \textbf{Conflict avalanche generation}\\
\\
Conflict avalanches refer to chains of related conflict events, with the relatedness determined by a statistical measure known as transfer entropy at an user-defined spatio-temporal scale. In this study, conflict avalanches were generated from the ACLED dataset covering the period 1997-2019, following the algorithm described in \cite{kushwahaDiscoveringMesoscaleChains2023a}. These conflict avalanches were clipped between the year 2000 and 2015 since that's the time period for which all the other datasets are available. One of the key advantages of this algorithm is its flexibility in allowing the selection of any spatio-temporal scale. Here, we set the spatial scale to approximately $b\approx66$\,km and the temporal scale to $a=30$\,days. Although our analysis is conducted at this scale, subsequent results indicate that the findings are robust across other scales as well (see Figure~\ref{appendixFig:PCA_SI_2}).
\\
\item \textbf{High dimensional conflict avalanche vectors}\\
\\
The conflict avalanche generation algorithm produces avalanches of varying sizes, measured by the number of constituent conflict events. We disregard avalanches consisting of a single conflict event, resulting in a final set of $5,659$ avalanches. Each avalanche $j$ is represented as a set 
\[
c_j = \{\vec{e}_i\}_j,
\]
where the size of $c_j$ corresponds to the number of events in the avalanche, and each element of $\vec{e}_i$ is a vector containing all available information about the corresponding conflict event $i$. Because the size of $c_j$ is variable and the components of $\vec{e}_i$ are in their raw continuous form (except for variables that are inherently discrete, such as the number of ethnic groups), this representation is highly complex and challenging to analyze given the available data. Consequently, it is necessary to compress this data while preserving the most relevant information encoded in $c_j$.
\\
\item \textbf{Compression via mean}\\
\\
An examination of the distribution of values for specific variables across conflict events within a given avalanche revealed that most values are concentrated around the mean. In fact, for almost all avalanches, the distribution is contained within one standard deviation of the mean, as illustrated in Figure~\ref{appendixFig:within_std}. Accordingly, each avalanche can be represented by a vector containing the mean values of each variable across its constituent conflict events. This approach compresses the variable-length avalanche vectors $c_j$ into fixed-length vectors corresponding to the total number of variables\footnote{In our study, we consider 22 variables in total, of which 19 are non-climatic. This mean-based compression is applied only to the non-climatic variables; climatic variables are addressed separately in the next section.}. Hence, each conflict avalanche is represented by a continuous vector of length 19 (see footnote 1), with each element corresponding to the mean value of the respective variable across the events comprising the avalanche.
\\
\item \textbf{Deviations vs absolutes}\\
\\
In the context of armed conflicts, deviations from a normative baseline are often of greater interest than absolute values. For example, when assessing the impact of economic prosperity, it is more informative to compare the onset and spread of conflicts between regions with relatively rich versus poor economies. Similarly, when considering environmental factors, the focus is on how \textit{changes} in climate affect conflict dynamics. With this perspective, we represent each conflict avalanche in terms of deviations from the norm.

For non-climatic variables, the aim is to evaluate how a given variable in a conflict-prone area of Africa deviates from the distribution observed across all such areas. For example, we often compare the GDP or population density of a particular area relative to other regions in Africa and draw connections between prevalence of conflicts in those areas. To quantify these deviations, the distribution of each variable across all conflict avalanches is partitioned into three bins using the 33rd and 66th percentile cutoffs. A variable value falling below the 33rd percentile is labeled as below median ($-1$), a value between the 33rd and 66th percentiles is labeled as at median ($0$), and a value above the 66th percentile is labeled as above median ($1$).

For climatic variables, we are not interested in deviations with respect to other areas but deviations from local history of a place. For example, how low or high were temperatures of a place with respect to the historical value when a conflict occurred at that place. Therefore, climatic deviations are measured relative to the local historical baseline rather than with respect to other areas. For each conflict event, the distribution of a climatic variable over the preceding 25 years at the specific location is partitioned into three bins using the 33rd and 66th percentiles. Values falling below the 33rd percentile are labeled as below median ($-1$), values between the 33rd and 66th percentiles as at median ($0$), and values above the 66th percentile as above median ($1$). For each conflict avalanche, a discrete value ($-1$, $0$, or $1$) is assigned by computing the mode of the labels across all conflict events within the avalanche.

Thus, each conflict avalanche is represented by a discrete vector of size 22, where each element takes one of the ternary values ($-1$, $0$, or $1$), corresponding to below, at, or above the median value, respectively. (Note: The numerical labels ($-1$, $0$, or $1$) here are simply symbolic.)
\\
\item \textbf{Bag of words}\\
\\
Despite the compression achieved so far, the data remain relatively complex, with a discrete state space of size $3^{22}$ and an entropy of approximately 17.7 bits. Consequently, the conflict avalanches occupy a discrete state space of size $\sim 10^5$, which, given the available number of avalanches, results in a sparsely populated high-dimensional variable space that may obscure meaningful similarities or differences between avalanches. To further simplify the representation, we employ a "bag-of-words" approach that compresses the data further but still captures the extremity (or median) of the variables within each variable category.

In this approach, for each variable category, we count the number of variables that fall below, at, or above the median range. This yields a vector of size three for each category, with each element corresponding to the count of variables classified as below, at, or above the median. For example, if an avalanche has below median values for temperature and precipitation and an at median value for NDVI, the climatic category vector for that avalanche would be $(2,1,0)$. Similarly, if an avalanche has an above median value for population count and an at median value for population density, the raw demographic category vector would be $(0,1,1)$.

By concatenating these bag-of-words representations across the six variable categories, each conflict avalanche is ultimately represented by a discrete vector of size 18. This final compressed vector is used in subsequent clustering of conflict avalanches using M4 (see~\ref{appendix:m4}).
\end{itemize}

\subsection{Step by step algorithm}\label{appendix:vector_algorithm}
Below is a concise, step-by-step algorithm for generating conflict avalanche vectors. The procedure is divided into two parts: one for climatic variables and one for all other (non-climatic) variables.\\

For climatic variables:\\
\begin{itemize}
    \item \textbf{Assign Variables to Conflict Events}: For each conflict event, attach the corresponding climatic variable values so that every event is fully characterized.
    \item \textbf{Generate Conflict Avalanches}: Group conflict events into conflict avalanches according to predefined spatial and temporal scales.
    \item \textbf{Construct Local Value Distributions}: For each geographical location, compile historical monthly values of each climatic variable (e.g., over the past 25 years) to establish a local baseline distribution.
    \item \textbf{Calculate Percentile Cutoffs}: For each variable at each location, compute the 33rd and 66th percentile thresholds. These thresholds divide the local distribution into three segments:
    \begin{itemize}
        \item Below Median: Values below the 33rd percentile.
        \item At Median: Values between the 33rd and 66th percentiles.
        \item Above Median: Values above the 66th percentile.
    \end{itemize}
    \item \textbf{Encode Variables for Conflict Events}: For each conflict event, determine in which percentile category its climatic variable values fall, and encode each variable accordingly (e.g., -1 for below median, 0 for at median, and 1 for above median).
    \item \textbf{Aggregate Variables for Conflict Avalanches}: Within each conflict avalanche, calculate the mode (i.e., the most frequent code) for each climatic variable across all constituent conflict events.
    \item \textbf{Construct Discrete Vectors for Conflict Avalanche}s: Combine the aggregated discrete values for all climatic variables into a single vector that represents each conflict avalanche.
\end{itemize}

\vspace{\baselineskip}

For non-climatic variables:\\
\begin{itemize}
    \item \textbf{Assign Variables to Conflict Events}: For each conflict event, assign the value of each non-climatic variable, ensuring that every event has complete data coverage.
    \item \textbf{Generate Conflict Avalanches}: Group conflict events into conflict avalanches based on the specified spatial and temporal scales.
    \item \textbf{Compute Avalanche-Level Variable Averages}: For each conflict avalanche, calculate the mean value of each non-climatic variable across all events within the avalanche.
    \item \textbf{Determine Percentile Cutoffs}: Across all conflict avalanches, determine the 33rd and 66th percentile thresholds for the distribution of each variable. These thresholds partition the distribution into three categories:
    \begin{itemize}
        \item Below median: Values below the 33rd percentile.
        \item At median: Values between the 33rd and 66th percentiles.
        \item Above median: Values above the 66th percentile.
    \end{itemize}
    \item \textbf{Encode Variable Values}: For each avalanche, encode each non-climatic variable’s average value based on the determined thresholds (e.g., -1, 0, or 1, corresponding to below, at, or above the median, respectively).

    \item \textbf{Construct Discrete Vectors}: Assemble the encoded values into a discrete vector for each conflict avalanche, with each entry corresponding to one of the non-climatic variables.
\end{itemize}

\clearpage

\section{SI figures}

\begin{figure}[!htbp]\centering
    \includegraphics[width=.35\linewidth]{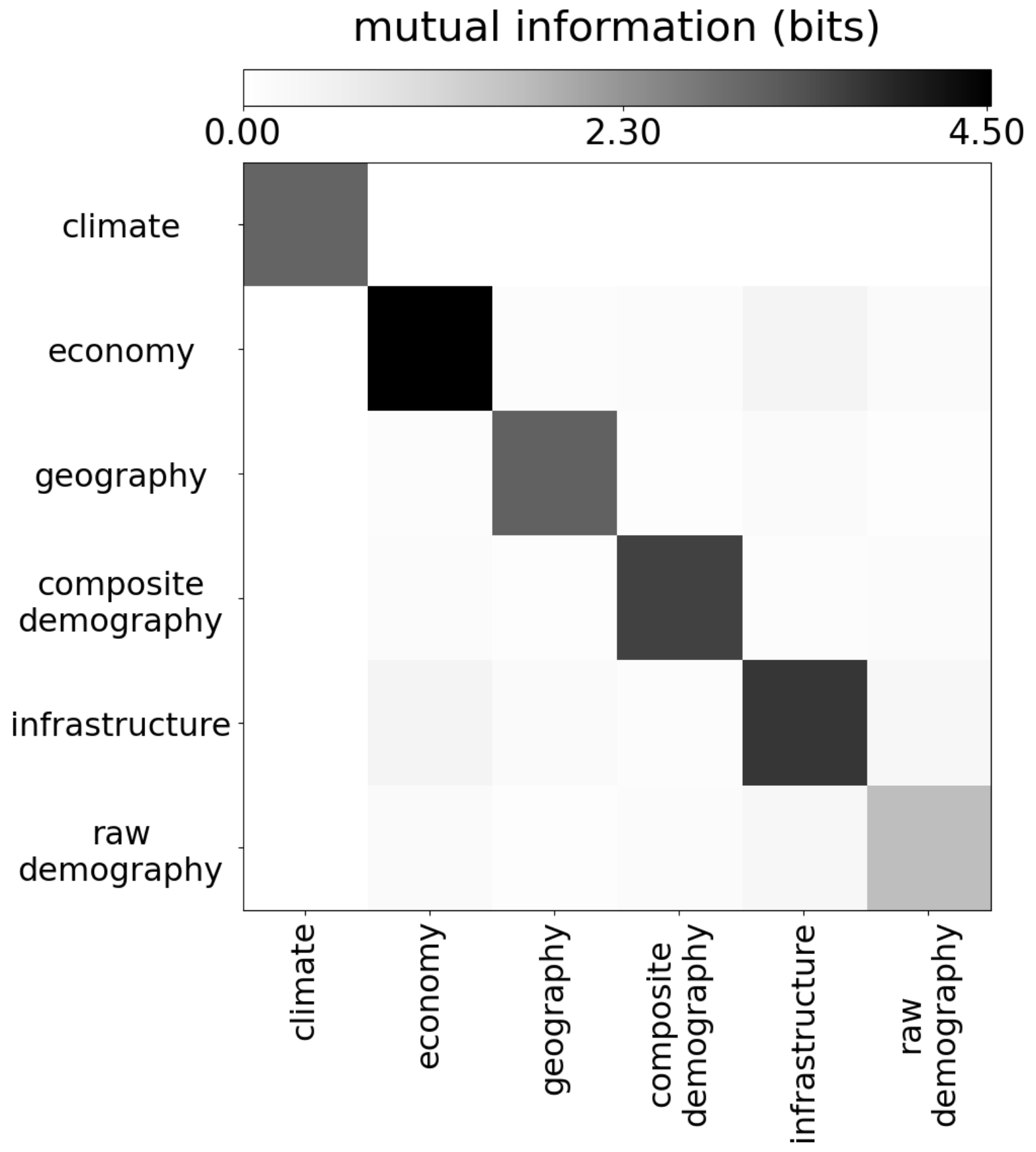}
    \caption{Mutual information between each variable category calculated after coarse-graining conflict vectors (see procedure in Appendix~\ref{appendix:conflict_vector}. The entropy shown along the diagonal summarizes the overall balance of variables in each category in terms of their entropy. The entropies are estimated using the NSB estimator \cite{nemenmanEntropyInference2002}.}
\label{appendixFig:MI_SI}
    \addcontentsline{toc}{subsection}{Figure \ref{appendixFig:MI_SI}: Mutual information between variable categories}
\end{figure}

\begin{figure*}[h]\centering
    \includegraphics[width=.5\linewidth, height=\textheight, keepaspectratio]{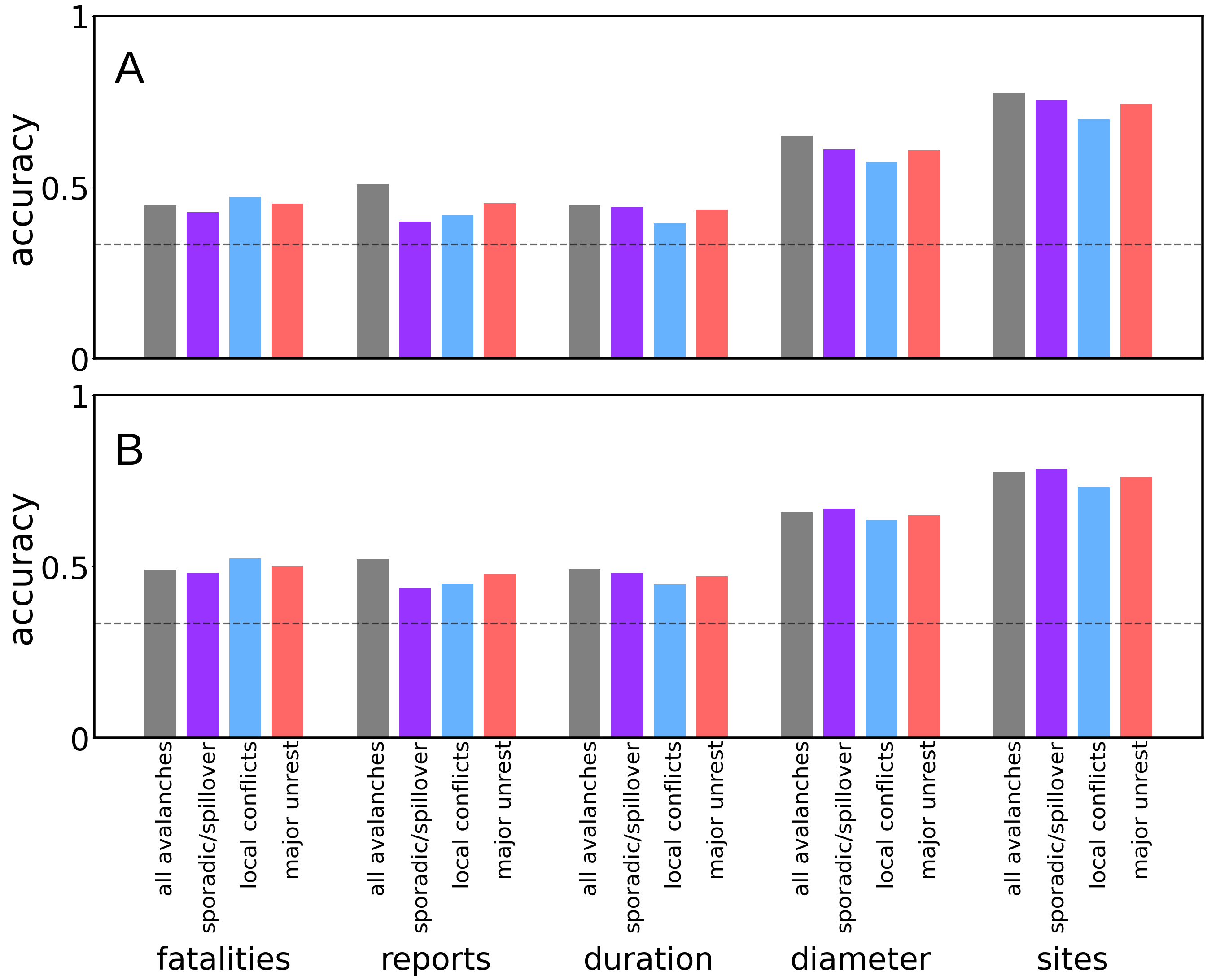}
    \caption{Accuracy of a random forest classifier in predicting whether a conflict avalanche’s features—namely, fatalities, number of reports, duration, diameter, and number of sites—fall below, at, or above their median values. Predictions were made under two conditions: training the model on the complete set of conflict avalanches and training it on three separate groups corresponding to the three conflict archetypes. The horizontal dashed line represents accuracy expected from a random classifier. A) for the case where we use the vectors where variable categories are in "bag of words" form (this plot is also shown in the main text in Figure~\ref{fig:prediction_fig}A) B) for the case where we use the vectors where all variables are considered individually.}\label{appendixFig:prediction}
    \addcontentsline{toc}{subsection}{Figure \ref{appendixFig:prediction}: Random forest model accuracy}
\end{figure*}

\begin{figure}[!htbp]\centering
    \includegraphics[width=\linewidth]{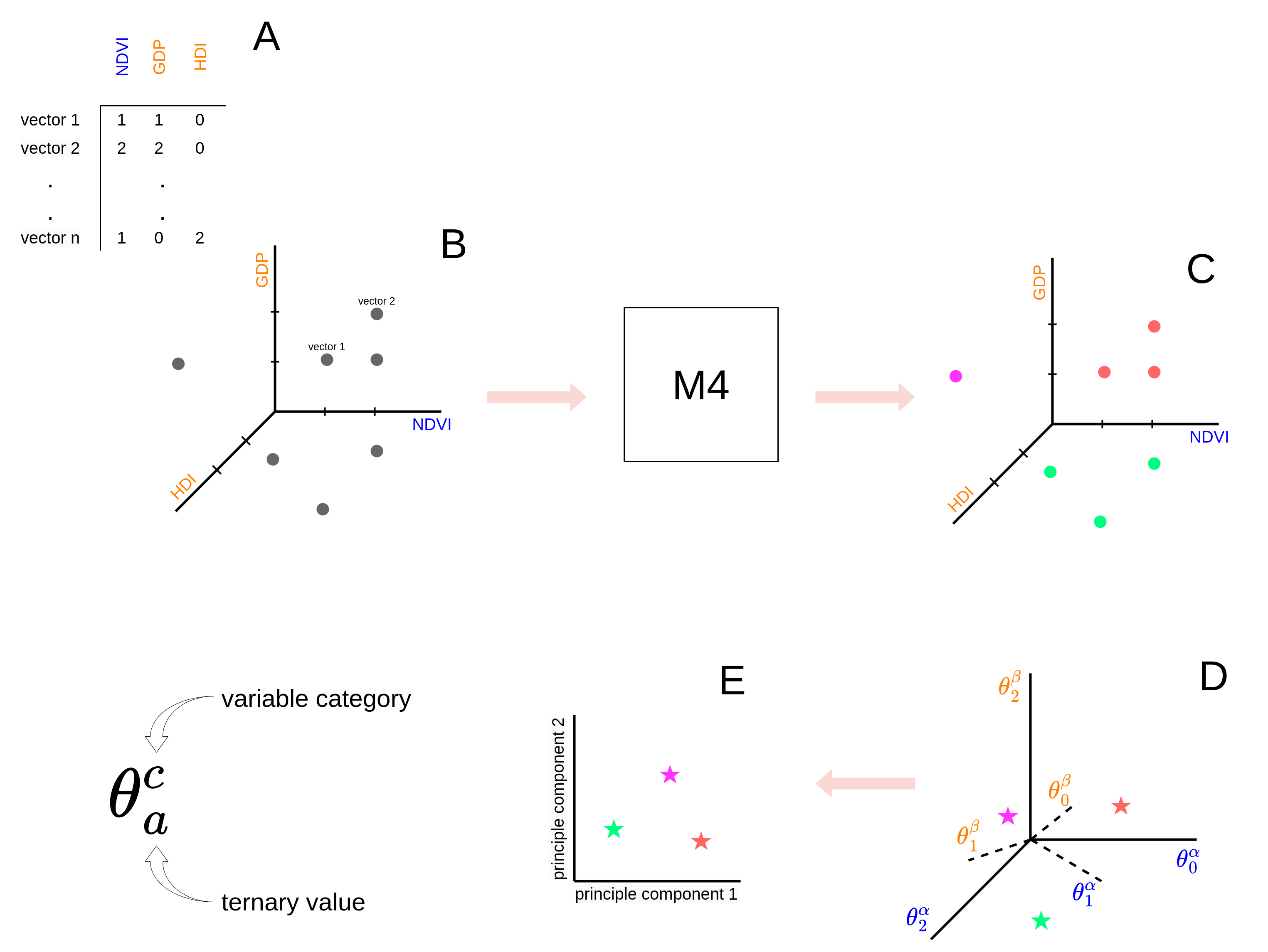}
    \caption{Schematic overview of the methodology. Conflict avalanche vectors are coarse‐grained (see Appendix~\ref{appendix:conflict_vector}) into discrete representations, where each element indicates whether a variable is below ($0$), at ($1$), or above ($2$) median (using $0,1,2$ instead of the standard $-1,0,1$ for easier visualization). A) A hypothesized example with $n$ avalanches is shown for three variables (NDVI, GDP, HDI) divided into climate (blue) and economic (orange) categories. B) These vectors are represented in a three‐dimensional discrete space. The M4 (see Appendix~\ref{appendix:m4}) clusters the vectors into $K$ clusters. C) shows an example where $K=3$ where avalanches belonging into same cluster is shown using same color. 
    Once we fit the M4, we get the parameter values which can be used to \textit{define}each cluster. D) Cluster parameters or centroids, defined in a $3 \times$(number of variable categories) parameter space (here, 6 dimensions for this hypothesized case) with $\vec{\theta}^\alpha = \left\{ \theta^\alpha_0,\theta^\alpha_1,\theta^\alpha_2 \right\}$ for climate and $\vec{\theta}^\beta = \left\{ \theta^\beta_0,\theta^\beta_1,\theta^\beta_2 \right\}$ for economic variables, are then projected onto the two dominant principal components via PCA shown in E). This is the plot readers see in the Figure~\ref{fig:proj} of the main text, for the actual data.}
\label{appendixFig:schematic}
	\addcontentsline{toc}{subsection}{Figure \ref{appendixFig:schematic}: Schematic overview of the methodology}
\end{figure}

\begin{figure}[!htbp]\centering
    \includegraphics[width=\linewidth]{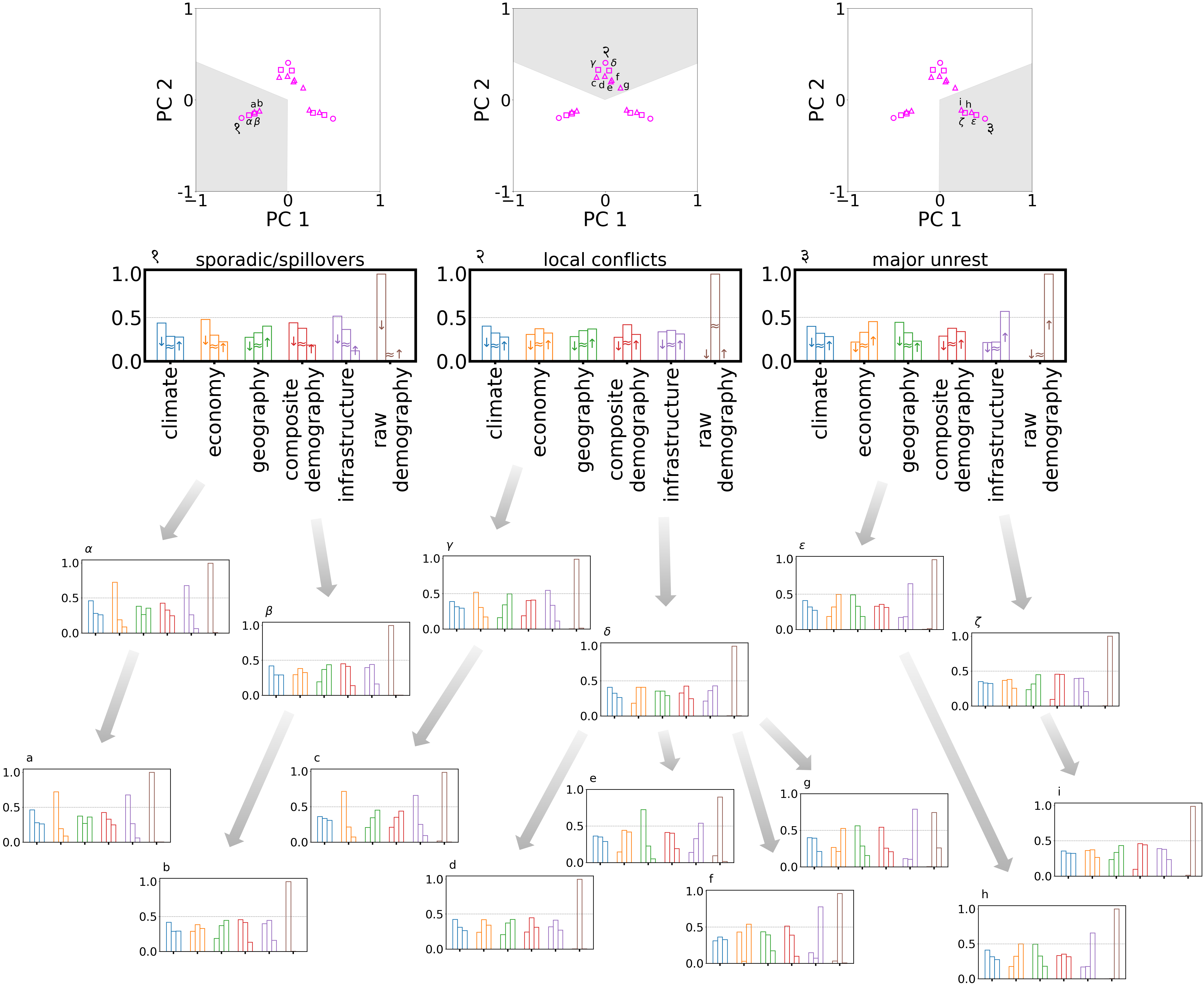}
    \caption{Cluster parameters at three different clustering tree depth $K$. The PCA biplots at the top serve as references for each panel below, indicating the corresponding cluster and the clustering level $K$. Clusters at $K=3,6,9$ are represented by circular, square and triangular markers respectively. Below, each panel shows the value of $\vec{\theta_j^c}$ corresponding to the labeled cluster, with three bars corresponding to each variable category. These bars denote the tendency of variables within each variable category, for that cluster, to fall below, at, or above the median, highlighted by $\downarrow, \approx$ and $\uparrow$ respectively. Panels {\dn 1}, {\dn 2}, and {\dn 3} depict the parameter values for sporadic/spillover conflicts, local conflicts, and major unrest, respectively. Major unrests, on average, tend to occur and spread in densely populated riparian zones or coastal plains with good infrastructure. Local conflicts show no discernible tendencies, generally appearing in areas of average population density. Sporadic/spillover conflicts are typically found in regions with low population density and poor infrastructure and economy. Panels labeled using Greek alphabets represent clusters obtained at $K=6$ while panels labeled using English alphabets represent clusters at $K=9$. Arrows illustrate the hierarchical division of clusters, demonstrating how clusters split as we increase $K$. To see the conflict avalanches that belong to each of these clusters see Figure~\ref{appendixFig:clusters}.}
\label{appendixFig:thetas}
\addcontentsline{toc}{subsection}{Figure \ref{appendixFig:thetas}: Clustering tree (cluster centroids)}
\end{figure}

\begin{figure}[!htbp]\centering
    \includegraphics[width=\linewidth]{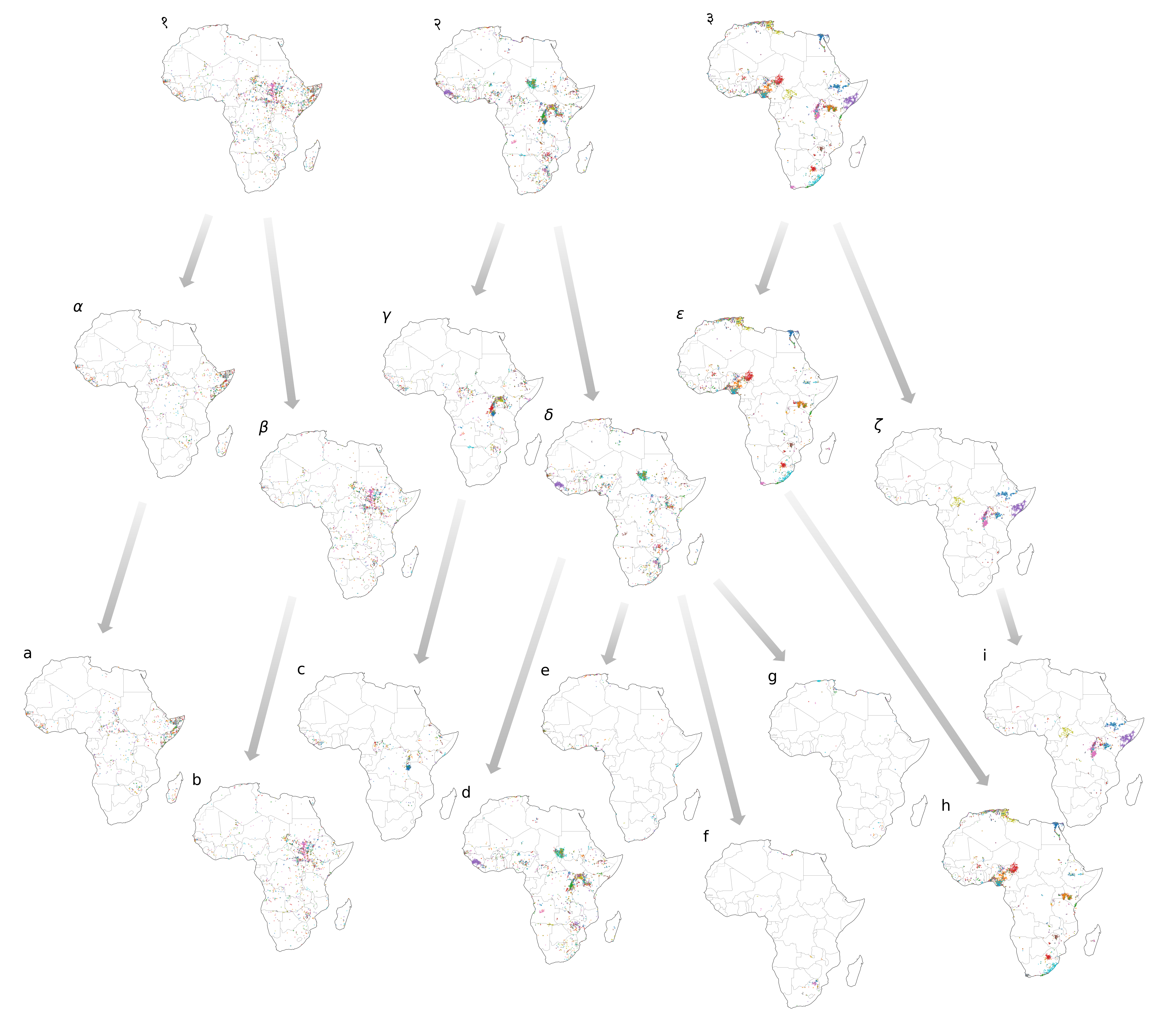}
    \caption{Conflict clusters shown at three clustering depths ($K$). Each map of Africa represents a distinct conflict cluster, containing conflict avalanches depicted using different colors. Panel labels correspond to clusters shown in PCA biplots of Figure~\ref{appendixFig:thetas}.
    Panels {\dn 1}, {\dn 2}, and {\dn 3} depict clusters for sporadic/spillover conflicts, local conflicts, and major unrest, respectively. Panels labeled using Greek alphabets represent clusters obtained at $K=6$ while panels labeled using English alphabets represent clusters at $K=9$. Arrows illustrate the hierarchical division of clusters, demonstrating how clusters split as we increase $K$. Parameter values for each of the clusters are shown in Figure~\ref{appendixFig:thetas}.}
\label{appendixFig:clusters}
\addcontentsline{toc}{subsection}{Figure \ref{appendixFig:clusters}: Clustering tree (conflict avalanches)}
\end{figure}

\begin{figure*}[h]\centering
    \includegraphics[width=\linewidth]{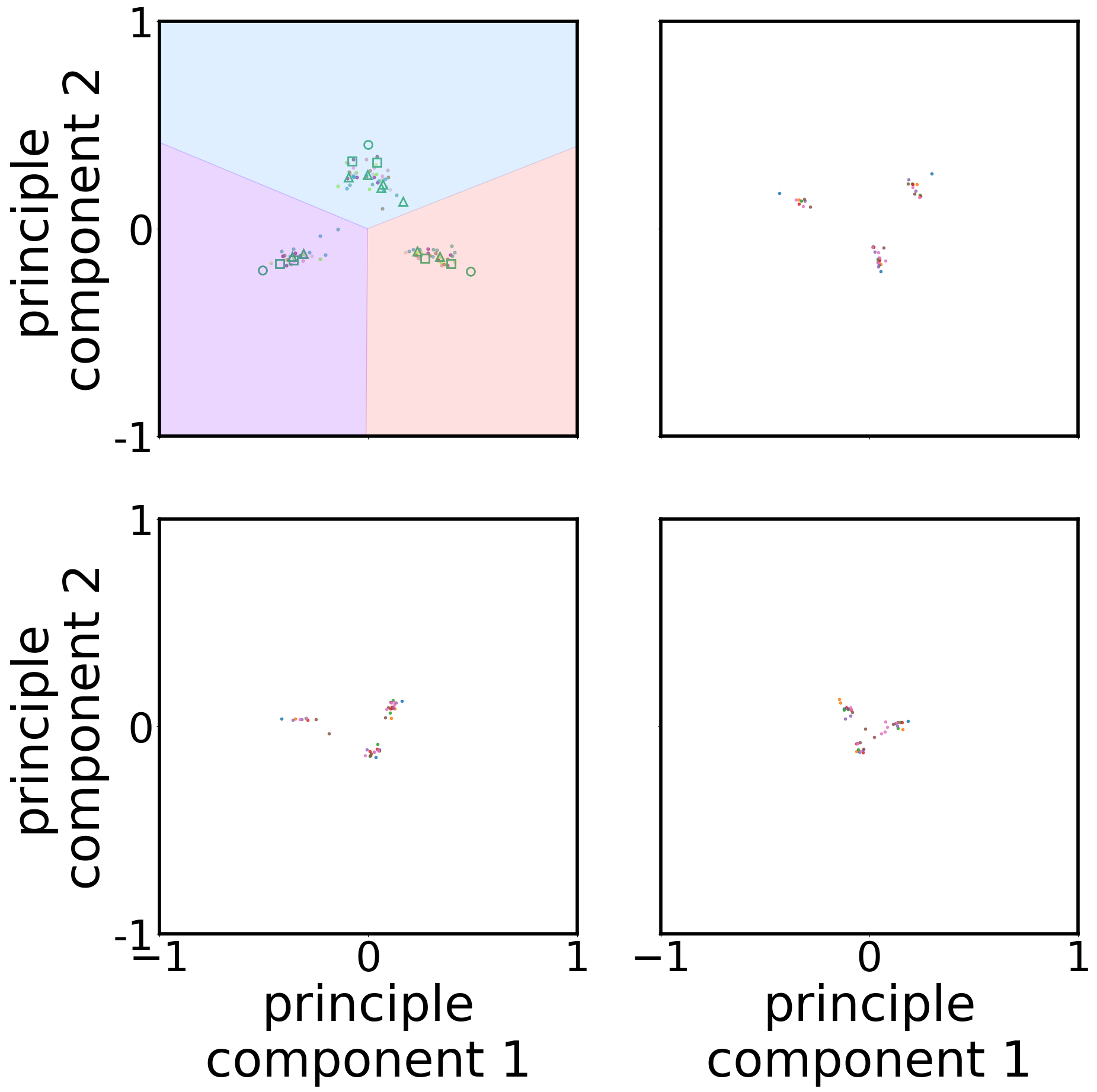}
    \caption{A) shows the cluster centroids projected onto the first two principle components for $2<K<16$. The hollow circular, square and triangular points corresponds to $K=3,6,9$ respectively. Next, we see cluster centroids projected onto the first two principle components for B) $L=4$ C) $L=5$ and D) $L=6$. B,C and D shows that the three archetypes are not an artifact of our coarse graining procedure where we compress the data into three bins at $L=3$.}\label{fig:PCA_SI}
    \addcontentsline{toc}{subsection}{Figure \ref{fig:PCA_SI}: Triangle of madness (varying $L$)}
\end{figure*}




\begin{figure}[h]\centering
    \includegraphics[width=.7\linewidth]{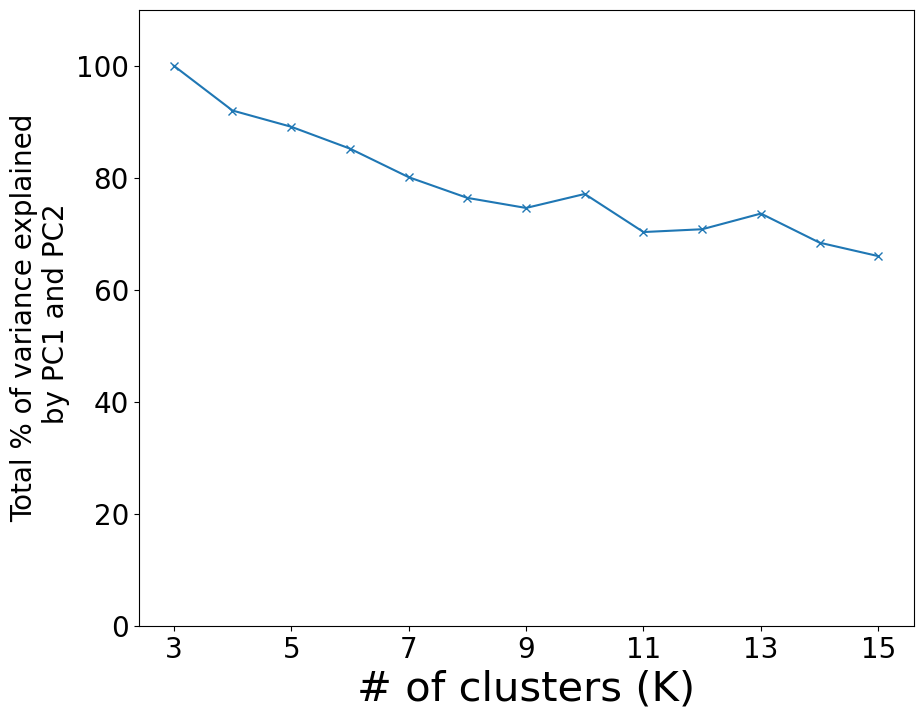}
    \caption{Variance in cluster centroids explained by the first two principle components.}\label{appendixFig:variance}
    \addcontentsline{toc}{subsection}{Figure \ref{appendixFig:variance}: Variance captured by principle components}
\end{figure}

\begin{figure}[h]\centering
    \includegraphics[width=\linewidth, height=\textheight, keepaspectratio]{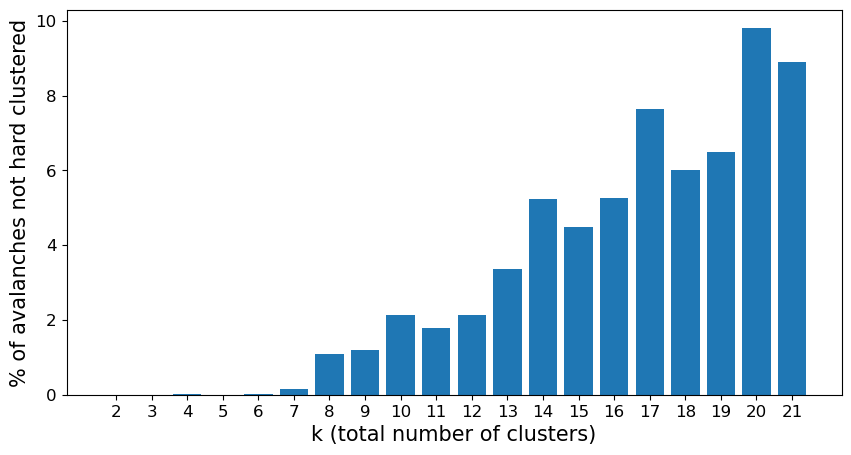}
    \caption{Percent of conflict avalanches that don't get hard clustered when the criteria for hard clustering is $\tau_{ij}>0.5$.}\label{appendixFig:not_hard_clustered}
    \addcontentsline{toc}{subsection}{Figure \ref{appendixFig:not_hard_clustered}: Percent of conflict avalanches not hard clustered}
\end{figure}

\begin{figure*}[h]
\centering
\includegraphics[width=\linewidth, height=\textheight, keepaspectratio]{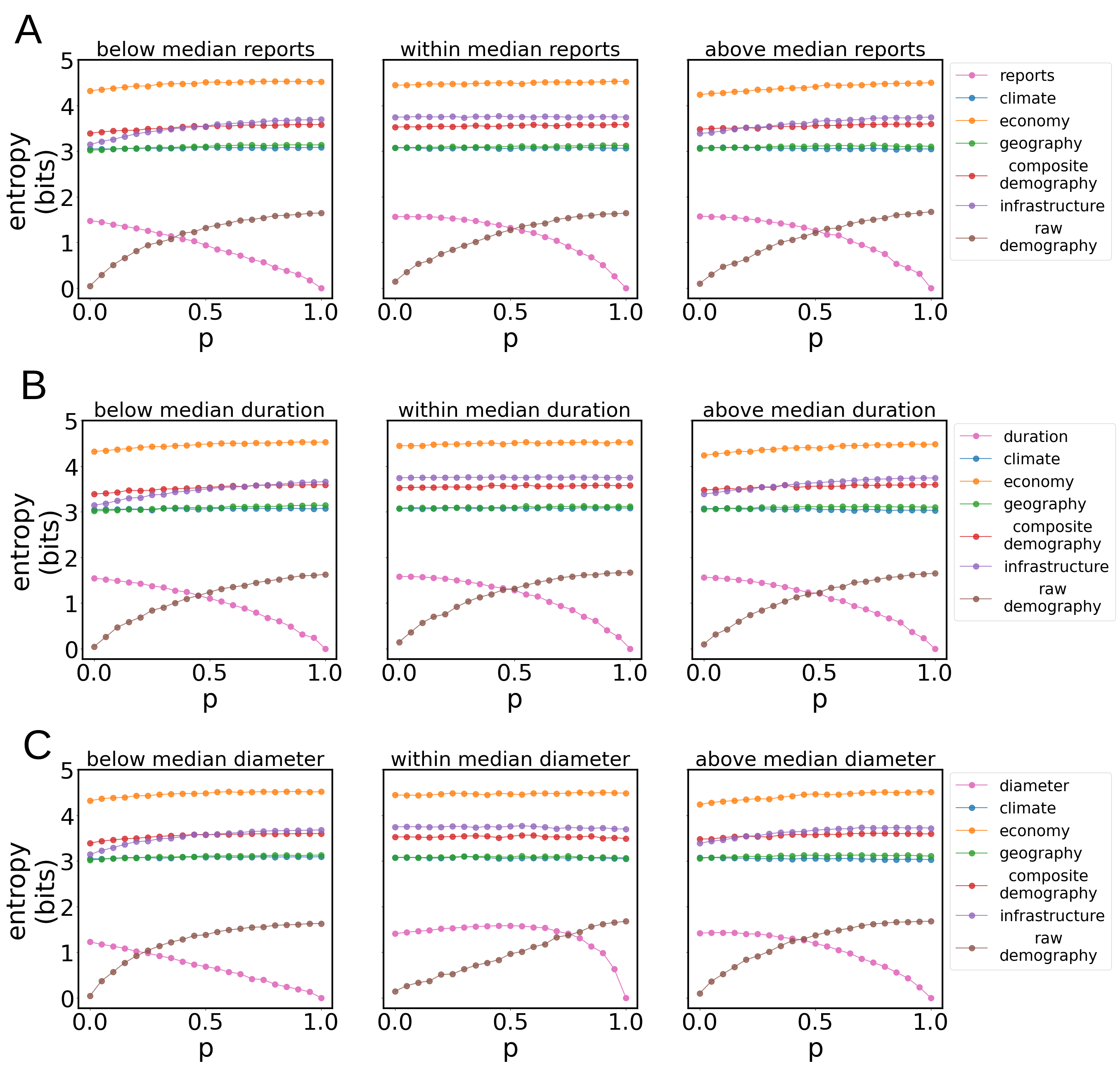}
\caption{The entropy trade-off between perfectly predicting a conflict’s archetype and perfectly predicting its intensity. Conflict intensity can be quantified by it's total A) fatalities, B) duration or C) diameter. The pink and brown curves indicate that increased certainty in predicting a conflict’s archetype (or its raw demographic value) corresponds to decreased certainty in predicting it's intensity, and vice versa. Here, $p$ is the probability with which any given conflict avalanche is placed into its cluster as given by M4 ($p=0$) or a model clustering based on it's intensity ($p=1$).}\label{fig:entropy_tradeoff}
        \addcontentsline{toc}{subsection}{Figure \ref{fig:entropy_tradeoff}: Entropy tradeoff for conflict intensity measures}
\end{figure*}

\begin{figure*}[h]\centering
    \includegraphics[width=\linewidth, height=\textheight, keepaspectratio]{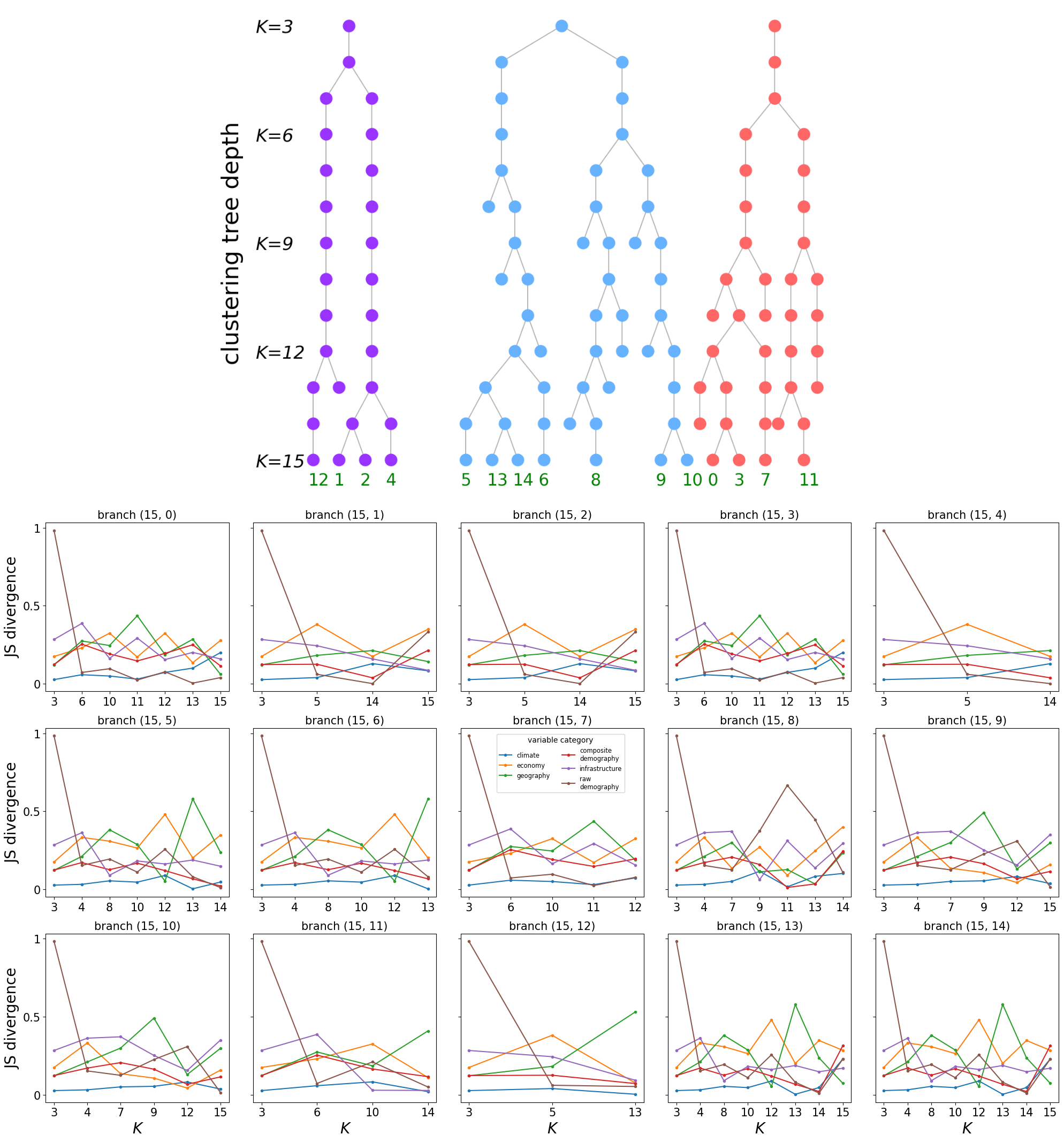}
    \caption{JS divergence between clusters at same $K$ of a given branch in the clustering tree, as a function of $K$. The title of each subplot represents the branch which can be identified by looking at the clustering tree shown above. Each branch is labeled as $(x,y)$ such that $x$ represents $K$ and $y$ represents cluster index (shown in green in the clustering tree).}\label{appendixFig:JS}
            \addcontentsline{toc}{subsection}{Figure \ref{appendixFig:JS}: JS divergence for each branch of the clustering tree}
\end{figure*}

\begin{figure*}[h]\centering
    \includegraphics[width=\linewidth, height=\textheight, keepaspectratio]{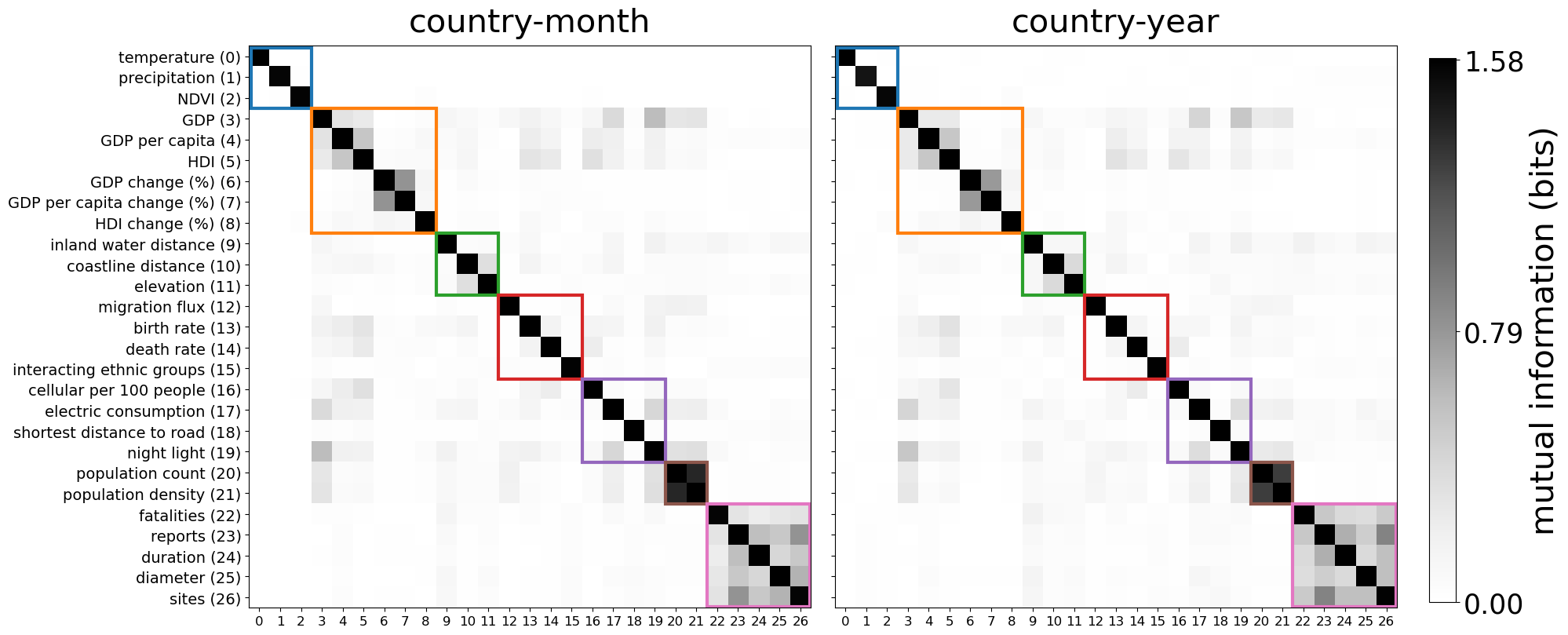}
    \caption{Mutual information matrix illustrating the mutual information values between pairs of variables.  Here, instead of aggregating conflict events by constructing conflict avalanches, we have aggregated them at the country-month and country-year levels. This approach is commonly employed in the literature when studying armed conflicts. Notably, the mutual information between variables in this case exhibits a similar pattern to that observed when using conflict avalanches (see Figure~\ref{fig:data}B in the main text). The conflict intensity variables—fatalities, reports, duration, diameter, and sites—exhibit higher mutual information in this case compared to conflict avalanches. This increase arises because, at this type of aggregation, these variables are typically proportional to the size of the country and the duration of the aggregation period.}\label{appendixFig:MI_country}
\addcontentsline{toc}{subsection}{Figure \ref{appendixFig:MI_country}: Mutual information for other conflict event groupings}
\end{figure*}

\clearpage  


\end{document}